\documentclass[aps,prapplied,reprint,superscriptaddress,longbibliography,floatfix]{revtex4-2}
\usepackage{amsmath,amssymb,bm,graphicx,booktabs,microtype}
\usepackage[colorlinks=true,linkcolor=blue,citecolor=blue,urlcolor=blue,filecolor=blue]{hyperref}
\hypersetup{pdftitle={Transferable Lindblad Variational Ansatz for Accelerated Steady-State Preparation},pdfauthor={Yun-Zhuo Fan and Dan-Bo Zhang},pdfsubject={Transferable steady-state preparation with compact modal diagnostics}}
\newcommand{\tr}{\operatorname{Tr}}
\newcommand{\ii}{\mathrm{i}}
\newcommand{\Lcal}{\mathcal L}
\newcommand{\Dcal}{\mathcal D}
\newcommand{\Hcal}{\mathcal H}
\newcommand{\Css}{\mathcal C_{\mathrm{s}}}
\newcommand{\Cloc}{\mathcal C_{\mathrm{loc}}}
\newcommand{\stead}{\mathrm{ss}}
\newcommand{\norm}[1]{\left\lVert#1\right\rVert}
\newcommand{\ket}[1]{\lvert#1\rangle}
\newcommand{\bra}[1]{\langle#1\rvert}
\newcommand{\Ucal}{\mathcal U}
\newcommand{\Bcal}{\mathcal B}
\begin{document}
\nocite{*}
\title{Transferable Lindblad Variational Ansatz for Accelerated Steady-State Preparation}
\author{Yun-Zhuo Fan}
\affiliation{Key Laboratory of Atomic and Subatomic Structure and Quantum Control (Ministry of Education), Guangdong Basic Research Center of Excellence for Structure and Fundamental Interactions of Matter, School of Physics, South China Normal University, Guangzhou 510006, China}
\author{Dan-Bo Zhang}
\email{dbzhang@m.scnu.edu.cn}
\affiliation{Key Laboratory of Atomic and Subatomic Structure and Quantum Control (Ministry of Education), Guangdong Basic Research Center of Excellence for Structure and Fundamental Interactions of Matter, School of Physics, South China Normal University, Guangzhou 510006, China}
\affiliation{Guangdong Provincial Key Laboratory of Quantum Engineering and Quantum Materials, Guangdong-Hong Kong Joint Laboratory of Quantum Matter, Frontier Research Institute for Physics, South China Normal University, Guangzhou 510006, China}
\date{November 2, 2026}
\begin{abstract}
Open quantum systems naturally relax toward steady states that underpin diverse nonequilibrium phenomena and quantum technologies. We independently control the coherent and dissipative constituents of a Lindblad generator to construct a variational product formula, termed the Lindblad variational ansatz (LVA), that accelerates steady-state preparation over natural relaxation through an optimized preparation path. The protocol consists of small-system learning, continuation to greater depth, and transfer to larger systems, followed when needed by reference-free post-training on quantum simulators. We demonstrate LVA in a bulk-dissipative Ising chain and a boundary-driven Heisenberg chain, where transferred protocols achieve matched preparation accuracies with fewer generator blocks than time-optimized first- and second-order Trotter formulas. The acceleration arises from coordinated endpoint preparation: earlier controls shape the dissipative-mode amplitudes so that the final dissipative step brings multiple components simultaneously close to their target values. Our results suggest a route to open-system quantum simulation that combines classical training on small systems with deployment and optional refinement on larger quantum devices.
\end{abstract}
\maketitle

\section{Introduction}\label{sec:introduction}
Open quantum systems relax toward steady states through the interplay of coherent interactions and dissipation. Such states can support nontrivial many-body correlations~\cite{verstraete2009,diehl2008,kastoryano2011}, nonequilibrium currents~\cite{prosen2011,znidaric2010,landi2022}, and collective order~\cite{diehl2010,lee2011,sieberer2013}. Preparing them is important both for understanding dissipative many-body physics and for quantum simulation~\cite{barreiro2011,schindler2013,weimer2021}. One route is to implement the target Lindblad dynamics directly. Slowly decaying modes can prolong this relaxation, while digital implementations require sufficiently fine steps to control errors from noncommuting generators~\cite{znidaric2015,macieszczak2016,trotter1959,strang1968,kliesch2011}. Although the generator fixes the steady state, its natural relaxation trajectory is only one path toward that state. Can we prepare the same state more efficiently without following that path?

Several approaches exploit this distinction. Reservoir engineering designs dissipative processes to stabilize desired states~\cite{verstraete2009,diehl2008,kraus2008}, while the quantum Mpemba effect shows that a suitable initial rotation can suppress a slow relaxation component~\cite{carollo2021,kochsiek2022}. Tensor-network and neural-network methods instead seek stationary-state representations without explicitly simulating relaxation~\cite{cui2015,hartmann2019,vicentini2019}. For quantum simulators that already implement the target generator, a complementary strategy is to retain its coherent and dissipative constituents while optimizing their ordering and durations. The resulting sequence need only prepare the correct endpoint; individual steps need not approximate natural evolution.

This direction connects naturally to physics-inspired variational quantum algorithms. Quantum annealing controls the balance between Hamiltonian terms, whereas the quantum approximate optimization algorithm (QAOA) turns alternating Hamiltonian evolutions into adjustable pulses~\cite{farhi2014,brady2021}. The Hamiltonian variational ansatz (HVA) likewise uses terms drawn from the physical model rather than generic circuit layers~\cite{wecker2015,wiersema2020,martin2022}. Such structured controls can also be reused: depth continuation provides effective initializations for deeper QAOA circuits, and selected HVA schedules transfer across system sizes~\cite{zhou2020,mbeng2019,mele2022}. For open systems, the analogous building blocks are the coherent and dissipative constituents of a Lindblad generator. A variational sequence that exploits their interplay extends the same physics-inspired approach to open systems, where transfer of learned controls across system sizes remains largely unexplored~\cite{endo2020,kamakari2022,liu2021}.

We introduce LVA as a generator-derived product formula whose coherent and dissipative constituents are independently controlled and jointly optimized for steady-state preparation. Unlike Trotterization, which approximates natural evolution, LVA optimizes the preparation endpoint. We show how coordinated endpoint preparation arises: earlier controls prepare dissipative-mode components that approach their target values together during the final dissipative step.

We also establish a transferable workflow. Controls are learned classically on a small system, extended in depth by continuation, and transferred to larger systems. Optional post-training minimizes a local Lindblad residual, providing a reference-free objective for quantum-simulator deployment. We demonstrate the workflow in an Ising chain with local bulk dissipation and a boundary-driven Heisenberg chain. In both demonstrations, transferred protocols reach matched infidelity thresholds with fewer generator blocks than time-optimized first- and second-order Trotter product formulas. Under common compiled generator templates, this reduction also lowers concatenated circuit depth.

The remainder of this paper is organized as follows. Section~\ref{sec:framework} introduces LVA and its learning objectives. Section~\ref{sec:learning} presents source learning and preparation benchmarks. Section~\ref{sec:diagnostics} examines preparation pathways, final-stage alignment, and parameter transfer. Section~\ref{sec:advantage} compares preparation resources after transfer and discusses circuit-depth implications. Section~\ref{sec:conclusion} concludes.

\section{Physical construction of the Lindblad variational ansatz}\label{sec:framework}
\subsection{Target dynamics and the preparation task}
A finite-dimensional Markovian open system obeys~\cite{lindblad1976,gks1976}
\begin{equation}
 \begin{aligned}[b]
 &\dot\rho=\Lcal_N(\rho)=-\ii[H_N,\rho]+
       \sum_\mu\gamma_\mu\Dcal[L_\mu](\rho),\\[2pt]
 &\Dcal[L](\rho)=L\rho L^\dagger-\tfrac12\{L^\dagger L,\rho\},
 \quad\gamma_\mu\geq0.
 \end{aligned}
 \label{eq:gkls}
\end{equation}
We consider a unique attractive steady state $\rho_{\stead}^{(N)}$ and a specified input $\rho_{0,N}$. For a diagonalizable generator, dual right and left eigenoperators satisfy $\tr(Q_k^\dagger R_j)=\delta_{kj}$, and
\begin{equation}
 \rho(t)-\rho_{\stead}=\sum_{k\ne0}c_k(0)e^{\lambda_k t}R_k,
 \quad c_k(0)=\tr[Q_k^\dagger(\rho_0-\rho_{\stead})].
 \label{eq:natural}
\end{equation}
Each mode decays at rate $\Gamma_k=-\operatorname{Re}\lambda_k$, so a populated slow mode can dominate the late-time deviation. LVA targets the same endpoint without requiring this particular trajectory. Preparing it from the specified input is distinct from stabilizing it under indefinite repetition of the learned sequence.

\subsection{Controlled layers and spatial sharing}
Decompose the target generator into available coherent and dissipative groups,
\begin{equation}
 \Lcal_N=\sum_{a=1}^{q}\Hcal_{a,N}+\sum_{b=1}^{s}\Dcal_{b,N},
 \quad \Hcal_{a,N}(\rho)=-\ii[H_{a,N},\rho].
 \label{eq:split}
\end{equation}
Each $\Dcal_{b,N}$ is itself a valid Lindblad generator. The depth-$P$ ansatz is
\begin{gather}
 \rho_{N,P}(\Theta)=\Phi_{P,N}\circ\cdots\circ\Phi_{1,N}(\rho_{0,N}),\label{eq:state}\\
 \Phi_{\ell,N}=
 e^{\tau_{s,\ell}\Dcal_{s,N}}\cdots e^{\tau_{1,\ell}\Dcal_{1,N}}
 e^{\theta_{q,\ell}\Hcal_{q,N}}\cdots e^{\theta_{1,\ell}\Hcal_{1,N}}.
 \label{eq:layer}
\end{gather}
The rightmost factor acts first. Dissipative exposures obey $\tau_{b,\ell}\geq0$; coherent parameters may have either sign when Hamiltonian reversal is available. Every factor is completely positive and trace preserving. Independent scheduling is the additional control resource relative to uninterrupted natural evolution~\cite{lloydviola2001}.

Coherent operations change the orientation and correlations of the state, whereas dissipation changes its spectrum. Optimizing their ordering and exposure permits preparation paths different from natural relaxation. Individual operations need not preserve the desired target, so the entire sequence is optimized jointly. Product-formula relaxation is a restricted choice of controls when ordering, parameter bounds, and sufficient depth agree; independent parameters release the requirement to approximate each short Lindblad step.

\begin{figure}[t]
 \centering\includegraphics[width=\columnwidth]{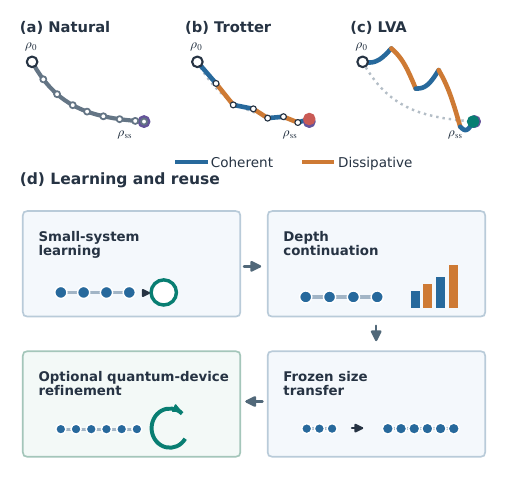}
 \caption{\label{fig:workflow}Preparation paths and learning workflow. (a) Natural relaxation approaches the target state. (b) Trotter steps alternate coherent (blue) and dissipative (orange) operations along the natural path. (c) Independently controlled LVA operations follow another path to a finite-protocol endpoint. The dotted curves are natural-relaxation references; colored endpoints are finite-protocol approximations, not fixed points. (d) Small-system learning, depth continuation, frozen size transfer, and optional quantum-device refinement. The last stage uses a local Lindblad residual; at fixed depth the shared-control parameter count is independent of system size. All paths are schematic.}
\end{figure}

Each parameter has a generator and layer index, but no site index. For $H_{a,N}=\sum_jh_{a,j}$, one angle controls the entire sum. The parameter count is $(q+s)P$, independent of $N$ at fixed depth. The models below have three controls per layer: two coherent angles and one dissipative exposure. Because each parameter retains the same generator-level meaning as $N$ changes, the learned schedule can be reused on a larger lattice. Figure~\ref{fig:workflow} summarizes the preparation and learning workflow.

\subsection{Implementing the dissipative blocks}
There are several routes to engineering or simulating dissipative dynamics, including measurement-based feedback~\cite{lloydviola2001}, reservoir engineering~\cite{weimer2010,barreiro2011}, and quantum-trajectory methods~\cite{dalibard1992,dum1992}. Here we describe unitary dilation as one general method for realizing the CPTP factors of LVA on a gate-based device~\cite{stinespring1955,wang2013channels,sweke2014,dive2015}. For a dissipative block $\mathcal E_\tau=e^{\tau\Dcal}$ at a chosen exposure $\tau\geq0$, a finite-dimensional Kraus representation is
\begin{equation}
 \mathcal E_\tau(\rho)=\sum_{j=0}^{r_K-1}K_j(\tau)\rho K_j^\dagger(\tau),
 \;\sum_{j=0}^{r_K-1}K_j^\dagger(\tau)K_j(\tau)=I.
 \label{eq:dissipativekraus}
\end{equation}
Introduce an auxiliary system $A$ with at least $r_K$ orthogonal states, initialized in $\ket0_A$. The Kraus completeness relation makes the map
\begin{equation}
 U_\tau(\ket\psi_S\otimes\ket0_A)
 =\sum_{j=0}^{r_K-1}K_j(\tau)\ket\psi_S\otimes\ket j_A
 \label{eq:dilationisometry}
\end{equation}
an isometry on the initialized subspace, which can be extended to a unitary on $S+A$. After applying this unitary, discarding the auxiliary realizes
\begin{equation}
 \mathcal E_\tau(\rho)
 =\tr_A\!\left[U_\tau(\rho\otimes\ket0\bra0_A)U_\tau^\dagger\right].
 \label{eq:dissipativedilation}
\end{equation}
A dissipative block is therefore implemented by initializing the auxiliary in $\ket0_A$, applying $U_\tau$ to the system and auxiliary, and tracing out the auxiliary. This realizes the channel without postselection. If the auxiliary is reused, it must be reset before the next block. The same dilation implements the dissipative factors in both LVA and Trotter circuits; their exposure schedules enter through the rotation angles. The required auxiliary dimension is set by the Kraus rank, while the gate cost is set by the chosen unitary synthesis.

\subsection{Learning objectives and physical accuracy}
For a small source system of size $N_s$, classical simulation provides the prepared state, while an independently computed steady state supplies a direct infidelity training objective. We use squared Uhlmann fidelity~\cite{jozsa1994,fuchs1999},
\begin{equation}
 F_N(\Theta)=\left[\tr\sqrt{\sqrt{\rho_{\stead}^{(N)}}\,
       \rho_{N,P}(\Theta)\sqrt{\rho_{\stead}^{(N)}}}\right]^2,
 \label{eq:fidelity}
\end{equation}
and minimize the source infidelity
\begin{equation}
 \Css(\Theta)=1-F_{N_s}(\Theta).
 \label{eq:sourcecost}
\end{equation}
Source training and schedule selection thus use the same accuracy measure as the preparation benchmarks. All source optimization uses only the small-system reference; no larger-system reference is needed to learn the transferable controls.

On larger quantum devices, the target density matrix may be unavailable. Lindblad stationarity supplies a reference-free post-training objective. Besides the global residual $\mathcal C_{\rm res}=\norm{\Lcal_N(\rho_{N,P})}_F^2$, we consider contiguous $r$-site regions $A_i$ and the local residual
\begin{equation}
 \Cloc^{(r)}(\Theta)=\frac{1}{N-r+1}\sum_i
    \norm{\tr_{\bar A_i}\Lcal_N[\rho_{N,P}(\Theta)]}_F^2.
 \label{eq:localcost}
\end{equation}
Its coefficients are expectations of $\Lcal_N^\dagger(O_{A_i})$, accessible through measurements on the finite neighborhood of $A_i$ for a finite-range generator. We assess each target-size preparation independently using the global infidelity $1-F_N(\Theta)$; the numerical reference is used for validation, not for post-training. We therefore separate source training, reference-free post-training, and the final fidelity assessment.

\begin{figure}[t]
 \centering\includegraphics[width=\columnwidth]{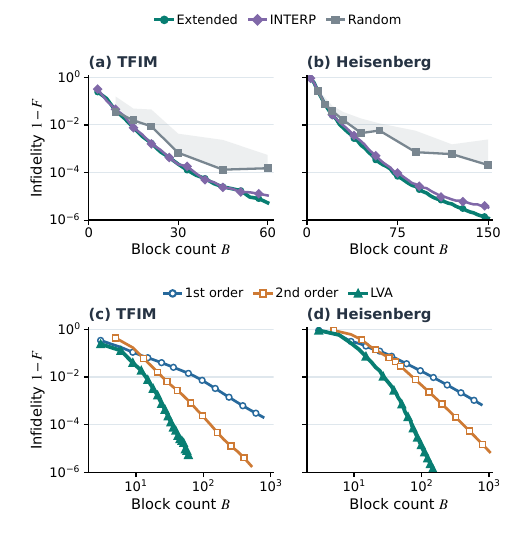}
 \caption{\label{fig:source}Source learning at $N_s=5$: TFIM (left) and Heisenberg (right). (a,b) Source infidelity versus the number $B=3P$ of generator operations: augmented depth continuation (Extended), ordinary INTERP, and the best of five random initializations (shading: full range). All minimize infidelity. (c,d) The learned source schedule (LVA) versus time-optimized first- and second-order Trotter product formulas. The same source-selected parameters are used for transfer. The methods are compared at selected depths; their total optimization budgets differ.}
\end{figure}

\section{Learning efficient preparation protocols}\label{sec:learning}
\subsection{Models and common controls}
We use Pauli matrices with $Z\ket0=\ket0$, $Z\ket1=-\ket1$, and $\sigma^-=(X-\ii Y)/2=\ket1\bra0$. Both open chains start from $\rho_{0,N}=(\ket1\bra1)^{\otimes N}$. The locally damped transverse-field Ising model (TFIM) has
\begin{equation}
 H_N=\sum_{j=1}^{N}X_j+\sum_{j=1}^{N-1}Z_jZ_{j+1},
 \quad \Dcal_N=\sum_{j=1}^{N}\Dcal[\sigma_j^-].
 \label{eq:tfim}
\end{equation}
The boundary-driven isotropic Heisenberg chain has
\begin{align}
 H_N&=\sum_{j=1}^{N-1}(X_jX_{j+1}+Y_jY_{j+1}+Z_jZ_{j+1}),\nonumber\\
 L_{L,+}&=\sigma_1^+,\quad L_{L,-}=\sqrt{0.2}\,\sigma_1^-,\nonumber\\
 L_{R,+}&=\sqrt{0.2}\,\sigma_N^+,\quad L_{R,-}=\sigma_N^-.
 \label{eq:xxx}
\end{align}
Here $\Dcal_N$ is the sum of the four boundary dissipators. Couplings and reference rates are unity and unchanged with $N$. The coherent groups are $\sum X$ and $\sum ZZ$ for TFIM, and $\sum(XX+YY)$ and $\sum ZZ$ for Heisenberg. Each layer is
\begin{equation}
 \Phi_{\ell,N}=e^{\tau_\ell\Dcal_N}
   e^{\theta_{2,\ell}\Hcal_{2,N}}e^{\theta_{1,\ell}\Hcal_{1,N}}.
 \label{eq:threelayer}
\end{equation}
For Heisenberg, a shared exposure scales the specified unequal pump and loss rates without changing their ratios. Each grouped operation is evaluated by its exact channel; any further decomposition into elementary hardware gates is a separate compilation task.

\subsection{Small-system learning and depth continuation}
The source size is $N_s=5$. Starting at shallow depth, INTERP resamples the learned coherent angles and dissipative exposures to initialize an additional layer~\cite{zhou2020,mbeng2019,mele2022}. The interpolated parameters are then optimized jointly. The deeper circuit inherits the order of physical processes while retaining freely optimized durations and angles.

We augment interpolation with identity-layer insertion, small perturbations, and source product-formula starting points. Identity insertion exactly preserves the preceding output before optimization; the other candidates allow neighboring stages to reorganize. For the source training used in Figs.~\ref{fig:source} and~\ref{fig:diagnosis}, each L-BFGS-B~\cite{byrd1995} call is capped at 1000 iterations, with earlier termination allowed when the convergence criteria are met. The parameter bounds are $\theta_{a,\ell}\in[-\pi,\pi]$ and $\tau_\ell\in[0,4]$. Augmented continuation, plain INTERP, and each random start all minimize the source infidelity in Eq.~\eqref{eq:sourcecost}; no target fidelity enters their training or selection. Source depths reach $P=20$ for TFIM and $P=50$ for Heisenberg. These runs are trained independently from depth one with the stated cap, rather than continued from a lower iteration budget. The modal, preimage, and transfer-path diagnostics use the same infidelity-trained source pool.

Figure~\ref{fig:source}(a,b) compares augmented continuation with plain INTERP and the best of five independent random starts at selected depths. Shading shows the spread among random starts. Augmented continuation obtains lower source losses at larger depths, with additional source evaluations used to identify stronger reusable schedules. This strategy follows the broader rationale of structured initialization and progressive variational training~\cite{grant2019,skolik2021,puig2025}.

\subsection{Source preparation and Trotter benchmarks}
We compare LVA with first- and second-order Trotter product formulas using the same generator groups and input. A first-order step is $e^{h\Dcal}e^{h\Hcal_2}e^{h\Hcal_1}$; the symmetric second-order step is
\begin{equation}
 S_2(h)=e^{\frac h2\Hcal_1}e^{\frac h2\Hcal_2}e^{h\Dcal}
     e^{\frac h2\Hcal_2}e^{\frac h2\Hcal_1}.
 \label{eq:strang}
\end{equation}
For each tested repetition count $m$, the total time $T=mh$ is optimized against the same reference fidelity. The alternative symmetric centering gives the same threshold conclusions. Each baseline is thus evaluated at its best tested relaxation time.

We count resources in generator blocks $B$, each denoting one complete coherent or dissipative exponential. The counts are $B=3P$ nominally for LVA, $3m$ for first order, and $4m+1$ for second order after merging adjacent identical half-steps. Zero-duration LVA factors remain in the nominal count. Pulse durations, compiled gate counts, and classical learning costs are distinct resources~\cite{preskill2018,cerezo2021review}.

Figure~\ref{fig:source}(c,d) shows error regimes in which the learned schedules reach a given fidelity with fewer generator blocks than the tested Trotter baselines. For example, the five-spin TFIM protocol at $P=7$ in this figure has infidelity $1.7\times10^{-3}$, compared with $1.6\times10^{-2}$ for the time-optimized second-order Trotter formula at the same block count. The gain comes from optimizing the preparation path, not from a more accurate approximation to each short Lindblad step. We next test whether this advantage is retained when the learned controls are transferred to larger systems.

\section{Preparation diagnostics and system-size transfer}\label{sec:diagnostics}
\begin{figure}[t]
 \centering\includegraphics[width=\columnwidth]{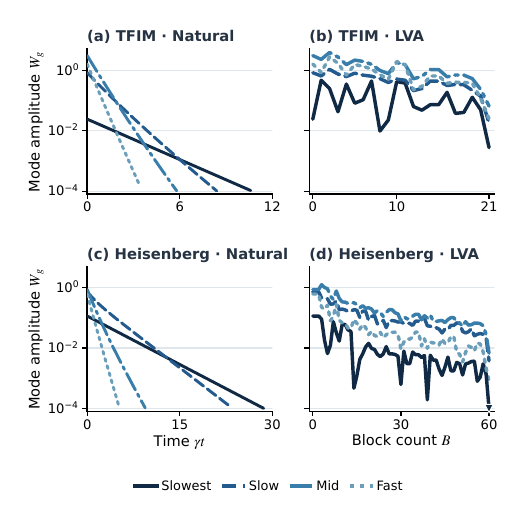}
 \caption{\label{fig:diagnosis}Mode-resolved preparation at $N=5$, arranged by model (rows) and protocol (columns). (a,c) Exact natural relaxation versus physical time. (b,d) LVA states after each generator operation, for TFIM ($P=7$) and Heisenberg ($P=20$). The four groups are distinguished by solid, dashed, dash-dotted, and dotted lines (dark to light shades for slow to fast decay); downward markers indicate values below the $10^{-4}$ display floor. The nonprobabilistic mode amplitudes diagnose preparation paths, not resource gains.}
\end{figure}

\subsection{Mode-resolved preparation paths}
To resolve how the learned paths differ from natural relaxation, expand $\rho_\ell-\rho_{\stead}$ in the target modes of Eq.~\eqref{eq:natural}. Normalize $\norm{R_k}_F=1$ and group the nonstationary coefficients by decay rate:
\begin{equation}
 W_g(\ell)=\left(\sum_{k\in g}|c_{\ell k}|^2\right)^{1/2},
 \quad c_{\ell k}=\tr[Q_k^\dagger(\rho_\ell-\rho_{\stead})].
 \label{eq:weights}
\end{equation}
Here $\Gamma_s$ is the slowest nonzero decay rate. The four groups are $\Gamma=\Gamma_s$, $\Gamma_s<\Gamma<a$, $a\leq\Gamma<b$, and $\Gamma\geq b$, with $(\Gamma_s,a,b)=(0.513,1.5,2.5)$ for TFIM and $(0.246,0.8,1.5)$ for Heisenberg.
A controlled layer obeys the affine update
\begin{equation}
 \mathbf c_\ell=M_\ell\mathbf c_{\ell-1}+\mathbf d_\ell.
 \label{eq:modeupdate}
\end{equation}
Here $M_\ell$ represents the layer acting on the nonstationary modes, while $\mathbf d_\ell$ is the modal projection of $\Phi_\ell(\rho_{\stead})-\rho_{\stead}$. Exact natural evolution instead has $M=\operatorname{diag}(e^{\lambda_k t})$ and $\mathbf d=0$. The affine update describes both mode mixing and intermediate displacement of the target. Matched-accuracy benchmarks establish the associated preparation advantage.

Figure~\ref{fig:diagnosis} compares exact natural relaxation with the infidelity-trained local-chain schedules: TFIM at $P=7$ and boundary-driven Heisenberg at $P=20$. Natural relaxation leaves a slow tail in both models. The TFIM LVA shows nonmonotonic reorganization followed by a joint endpoint decrease, whereas Heisenberg exhibits more distributed suppression with revivals. The $W_g$ are coefficient norms in a generally nonorthogonal basis, rather than populations or fidelity errors.

\subsection{Last-layer alignment with the target}
The final dissipative block resolves the endpoint effect in the spectral basis of the dissipator. For the single dissipative generator group used in our models, write $\Bcal_P=e^{\tau_P\Dcal}$ and denote its physical input by $\rho_{\rm pre}=\Ucal_P(\rho_{P-1})$. The target $\rho_{\stead}$ is stationary under the full $\Lcal=\Hcal+\Dcal$, but generally not under $\Dcal$ alone. For these diagonalizable dissipators with real decay rates, let $\mathcal P_g$ be the algebraic spectral projector satisfying $\Dcal\mathcal P_g=-\Gamma_g\mathcal P_g$, and define
\begin{equation}
 C_g=\mathcal P_g\rho_{\stead},\quad A_g=\mathcal P_g\rho_{\rm pre}.
 \label{eq:preimagecomponents}
\end{equation}
These are dissipative spectral groups, distinct from the full-Liouvillian mode groups in Fig.~\ref{fig:diagnosis}. Stationarity gives $\Gamma_gC_g=\mathcal P_g\Hcal(\rho_{\stead})$: even groups with $\Gamma_g>0$ can have nonzero target components that must be prepared rather than eliminated.

During the final block, each group evolves independently as
\begin{equation}
 \mathcal P_g\rho_P=e^{-\Gamma_g\tau_P}A_g,\quad A_g\simeq e^{\Gamma_g\tau_P}C_g.
 \label{eq:preimagematching}
\end{equation}
The second relation gives the endpoint-matching condition. We call this coordinated endpoint preparation: earlier coherent and dissipative blocks prepare the input components in direction and amplitude so that the final block brings them close to their target values at a common time. Within the circuit's reachable set, this requires a physical approximation to the dissipative preimage $e^{-\tau_P\Dcal}(\rho_{\stead})$. This inverse defines the desired preimage; it is not an operation applied in the protocol.

For $\norm{C_g}_F>0$, define
\begin{equation}
 r_g(t)=\frac{\norm{e^{-\Gamma_gt}A_g-C_g}_F}{\norm{C_g}_F},\quad t_g^*=\operatorname*{arg\,min}_{t\geq0}r_g(t).
 \label{eq:groupmismatch}
\end{equation}
At the optimized endpoint, each group must match the target in both amplitude and direction at a common time near $\tau_P$. Zero-rate groups are prepared by the preceding controls because the terminal dissipator leaves them unchanged. 

\begin{figure}[t]
 \centering\includegraphics[width=\columnwidth]{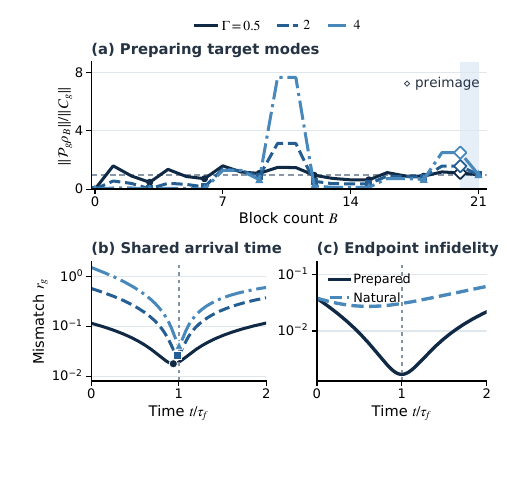}
 \caption{\label{fig:preimage}Dissipative preimage preparation in TFIM ($N=5$, $P=7$). (a) Target-normalized spectral amplitudes after each generator operation $B$. The horizontal dashed line is the target amplitude, open diamonds indicate required preimage amplitudes $e^{\Gamma_g\tau_f}$, and shading denotes the last dissipative operation. (b) Mismatch for the fixed pre-dissipation state as its final damping duration varies. Markers locate the group-specific minima. (c) Infidelity after the same damping channel acts on the learned input or a natural-relaxation input matched in initial infidelity. In (b,c), the vertical line denotes the learned terminal duration $\tau_f=\tau_P$. All trajectories are computed from the saved source controls.}
\end{figure}

Figure~\ref{fig:preimage} tests this picture using the same infidelity-trained five-spin TFIM protocol at $P=7$ as Fig.~\ref{fig:diagnosis}. Its initial state is a fixed point of $\Dcal$, so the selected positive-rate components start at zero although their target values are nonzero. Earlier blocks establish these components before the last dissipative operation. At the trained $\tau_P\simeq0.23$, the $\Gamma_g=0.5,2,4$ groups have $t_g^*/\tau_P\simeq0.94,0.99,1.00$, respectively. These rates span the slow, intermediate, and fast parts of the dissipative spectrum with appreciable target components. Despite different decay rates, their mismatches reach minima near the same finite time. The block checkpoints are samples of continuous dissipative evolution.

The input-state comparison in Fig.~\ref{fig:preimage}(c) isolates the role of this preparation. An input on the exact natural-relaxation trajectory is chosen to have the same pre-dissipation infidelity, $0.038$, as $\rho_{\rm pre}$. Applying the identical $e^{\tau_P\Dcal}$ gives endpoint infidelities $0.031$ and $0.0017$, respectively. The learned input is therefore better matched to the final channel despite equal initial infidelity. This approximately 18-fold difference in endpoint error is distinct from the resource gain measured below.

\begin{figure}[t]
 \centering\includegraphics[width=\columnwidth]{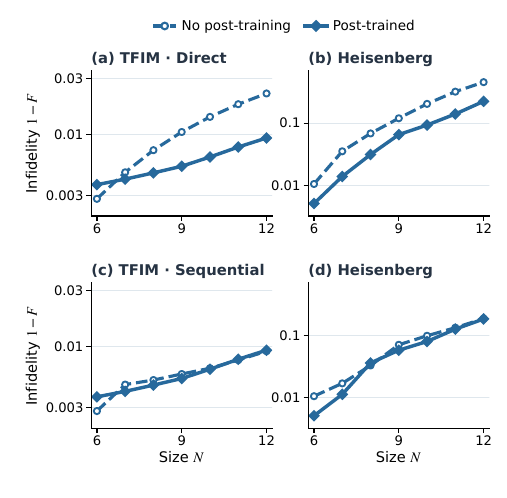}
 \caption{\label{fig:transfer}Transfer from $N_s=5$ to larger systems. Columns show TFIM ($P=7$) and Heisenberg ($P=20$); rows show direct (a,b) and sequential (c,d) transfer. ``No post-training'' means that transferred controls are evaluated without further optimization at the displayed size $N$; sequential controls may already have been optimized at smaller sizes. ``Post-trained'' means that those controls are refined at $N$ using two-site (TFIM) or four-site (Heisenberg) local Lindblad residuals.}
\end{figure}

\subsection{Parameter reuse and post-training}
Spatially shared parameters carry the learned protocol to larger systems. Optional target-size post-training adjusts these controls by minimizing the local Lindblad residual. We consider two ways to reuse them. In direct transfer, each larger system starts from the same source controls. In sequential transfer, it starts from the final controls refined at the preceding size along a chosen path. The sequential path may traverse every intermediate size or skip some sizes, with optional refinement at the final target. Couplings, input convention, generator ordering, and parameter meanings are unchanged throughout.

Figure~\ref{fig:transfer} compares direct and unit-step sequential transfer for the same infidelity-trained source schedules, with $P=7$ for TFIM and $P=20$ for Heisenberg. Refinement minimizes Eq.~\eqref{eq:localcost}, using $r=2$ and $r=4$, respectively, with a 1000-iteration cap per L-BFGS-B call and convergence-based termination allowed. The final local-cost optimizer output is retained; target infidelity does not select an intermediate iterate or train the controls. These numerical tests use exact expectation values.

The figure label ``No post-training'' applies only to the displayed target size. For direct transfer it means using the controls learned at $N_s=5$; for sequential transfer to $N$, it means using controls already post-trained through $N-1$.

For TFIM, direct-transfer infidelity grows gradually with system size, and two-site residual refinement improves it at $N=7,\ldots,12$. At $N=6$, however, refinement changes infidelity from $0.0028$ to $0.0037$ despite reducing the local cost. This difference reflects the distinct objectives of local stationarity and global state fidelity. At $N=12$, refinement reduces infidelity from $0.022$ to $0.0093$. Sequential transfer reaches $0.0091$ before refinement at that size, already below the refined direct-transfer result.

For Heisenberg, direct-transfer infidelity grows more strongly with system size, but four-site refinement improves it at every tested size. The same control coordinates therefore remain useful after transfer in this boundary-driven example, although their accuracy depends more strongly on system size. At $N=12$, refinement reduces infidelity from $0.46$ to $0.22$, whereas sequential transfer already reaches $0.19$ before target refinement.

For both models at $N=12$, sequential transfer before target refinement gives a lower infidelity than direct transfer after target refinement, even though the sequential controls have been optimized only at smaller sizes. Intermediate-size training thus provides a more accurate initialization for the largest target and shifts part of its optimization effort to smaller systems.

\section{Transfer of the preparation advantage}\label{sec:advantage}
\subsection{Block advantage at matched accuracy}
We quantify preparation efficiency through the smallest tested block counts that meet a common infidelity threshold. Define the sampled block gain over symmetric second order by
\begin{equation}
 \widehat G_B^{(2)}(N,\epsilon)=
 \frac{\widehat B_2(N,\epsilon)}{\widehat B_{\mathrm{LVA}}(N,\epsilon)},
 \label{eq:gain}
\end{equation}
where each budget is the smallest evaluated block count attaining $1-F_N\leq\epsilon$. For sequential transfer, the ``No post-training'' controls may already include optimization at smaller sizes; target-size post-training is assessed separately. Target reference states are used to optimize the baseline's total time and assess final fidelity, but not to select source schedules or train target LVA controls. The comparison uses the source-infidelity training and local-residual refinement described for Fig.~\ref{fig:transfer}, with a 1000-iteration cap per optimizer call. 

\begin{figure}[t]
 \centering\includegraphics[width=\columnwidth]{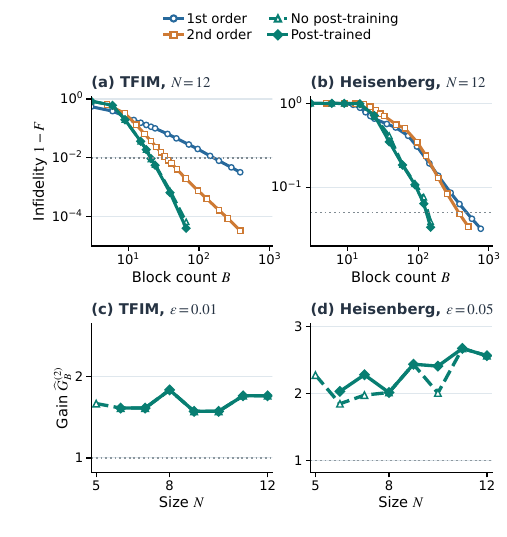}
 \caption{\label{fig:gain}Transferable preparation advantage under sequential size transfer. (a,b) At $N=12$, LVA without or with target-size post-training compared with time-optimized first- and second-order Trotter product formulas. (c,d) Sampled operation-count gain over second order for $N=5,\ldots,12$, at infidelity thresholds $0.01$ (TFIM) and $0.05$ (Heisenberg). Open dashed triangles denote no post-training at the displayed size; solid diamonds denote post-training there. The nominal LVA operation count is $B=3P$. Each ratio uses the smallest successful evaluated budget, not a certified optimum.}
\end{figure}

Figure~\ref{fig:gain} compares sequentially transferred preparations with first- and second-order Trotter product formulas. At $N=12$, TFIM reaches the one-percent infidelity threshold with $B=21$ both before and after target refinement, compared with the smallest successful tested second-order budget $B=37$. The gain is $37/21\simeq1.76$. Across $N=5,\ldots,12$, TFIM gains range from about $1.57$ to $1.83$.

Heisenberg provides a more demanding test. At $N=12$, sequential transfer reaches the five-percent infidelity threshold with $B=150$ before refinement; four-site refinement at the same depth reduces infidelity from $0.039$ to $0.032$. The second-order Trotter baseline requires $B=385$, giving a gain of $385/150\simeq2.57$ at either stage. Refinement can also lower the threshold budget: at $N=6,7,10$, the counts decrease from $33$ to $30$, $45$ to $39$, and $108$ to $90$, respectively. The sampled gains exceed unity across all tested sizes, with a nonmonotonic dependence on $N$. The two models use different thresholds because their accessible global fidelities differ; every resource ratio compares a common threshold within its model.

\subsection{Implications for hardware circuit depth}
The block advantage translates into circuit-depth savings when the generator subcircuits share a compilation and scheduling scheme. At fixed $N$ and synthesis accuracy, let $d_1,d_2,d_D$ be the sequential depths of the two coherent templates and the dissipative template, including auxiliary preparation/reset stages in $d_D$. If their depths are independent of the control angles over the range used, concatenating the templates gives
\begin{align}
 D_{\rm LVA}&=P(d_1+d_2+d_D),\nonumber\\
 D_1&=m(d_1+d_2+d_D),\nonumber\\
 D_2&=(m+1)d_1+2m d_2+m d_D.
 \label{eq:hardwaredepth}
\end{align}
The second-order count merges adjacent half-steps. LVA and first-order Trotterization use identical block patterns, so fewer repetitions require fewer executions of each compiled subcircuit. For the local damping channels used here, the dilation circuits have a fixed gate structure: changing the exposure changes the rotation angles rather than the number of channel repetitions.

The matched-accuracy examples at $N=12$ also imply smaller depth than the stated symmetric baseline under this compilation model. TFIM uses $P=7$ versus $m=9$; hence $D_2-D_{\rm LVA}=3d_1+11d_2+2d_D>0$. Heisenberg uses $P=50$ versus $m=96$, giving $D_2-D_{\rm LVA}=47d_1+142d_2+46d_D>0$. These expressions use the evaluated endpoint budgets and leave the individual compiled depths explicit.

For TFIM, the coherent groups consist of parallel on-site rotations and commuting nearest-neighbor $ZZ$ gates, which can be scheduled on even and odd bonds. Local dilations likewise have depth independent of $N$ when sitewise auxiliaries and parallel reset are available. The Heisenberg $XX+YY$ group contains noncommuting overlapping bonds; its exact numerical exponential requires a suitable hardware synthesis. With common compiled generator templates, the smaller matched-accuracy repetition budgets in Fig.~\ref{fig:gain} yield smaller concatenated circuit depths. 

In the boundary-driven Heisenberg chain, the coherent $XX+YY$ group transports excitations injected at the two ends into the bulk. The transfer results therefore reflect the combined action of boundary pumping and coherent transport.

\section{Conclusions}\label{sec:conclusion}
We have developed the Lindblad variational ansatz, a generator-derived product formula that independently controls the coherent and dissipative constituents of a target Lindbladian and optimizes their sequence for steady-state preparation. In bulk-damped Ising and boundary-driven Heisenberg demonstrations, LVA has achieved matched infidelity thresholds using fewer generator blocks than time-optimized first- and second-order Trotter formulas.

We have identified how preceding controls prepare dissipative-mode components for coordinated final-stage alignment with the target. We have also demonstrated that controls learned classically on small systems retain their preparation advantage after depth continuation and transfer to larger systems. Local Lindblad residuals have provided a reference-free objective for optional post-training on quantum simulators. These results have established a route from classical training on small systems to deployment and further optimization on larger quantum devices.

\begin{acknowledgments}
This work was supported by the National Natural Science Foundation of China (Grants No. 12375013, No. 12547109), and Guangdong Provincial Quantum Science Strategic Initiative (Grant No. GDZX2503008).
\end{acknowledgments}

\section*{Data availability}
The data and code that support the findings of this article are openly available~\cite{lva_data_code}.

\bibliographystyle{lva-preserve}
\bibliography{bibliography/main}
\end{document}


\nocite{*}
\title{Supplemental Material\texorpdfstring{\\}{: }Transferable Lindblad Variational Ansatz for Accelerated Steady-State Preparation}
\author{Yun-Zhuo Fan}
\affiliation{Key Laboratory of Atomic and Subatomic Structure and Quantum Control (Ministry of Education), Guangdong Basic Research Center of Excellence for Structure and Fundamental Interactions of Matter, School of Physics, South China Normal University, Guangzhou 510006, China}
\author{Dan-Bo Zhang}
\email{dbzhang@m.scnu.edu.cn}
\affiliation{Key Laboratory of Atomic and Subatomic Structure and Quantum Control (Ministry of Education), Guangdong Basic Research Center of Excellence for Structure and Fundamental Interactions of Matter, School of Physics, South China Normal University, Guangzhou 510006, China}
\affiliation{Guangdong Provincial Key Laboratory of Quantum Engineering and Quantum Materials, Guangdong-Hong Kong Joint Laboratory of Quantum Matter, Frontier Research Institute for Physics, South China Normal University, Guangzhou 510006, China}
\maketitle

\section{Models and independently computed reference states}
\label{sec:models}
The calculations use the two open chains defined in the article, with $N_s=5$. At the source size, classical optimization minimizes the infidelity $\mathcal C_s(\Theta)=1-F(\rho_{N_s,P}(\Theta),\rho_{\rm ref}^{(N_s)})$; target refinement instead minimizes the local stationarity residual. The source-infidelity transfer and resource tests extend through $N=12$. All rates, couplings, generator ordering, and initial-state conventions remain unchanged under transfer. We set $Z\ket0=\ket0$, $Z\ket1=-\ket1$, and $\sigma^-=\ket1\bra0$. The initial state is $(\ket1\bra1)^{\otimes N}$. In the TFIM, $H_1=\sum_jX_j$, $H_2=\sum_jZ_jZ_{j+1}$, and $\Dcal_N=\sum_j\Dcal[\sigma_j^-]$. In Heisenberg, $H_1=\sum_j(X_jX_{j+1}+Y_jY_{j+1})$, $H_2=\sum_jZ_jZ_{j+1}$, and the pump/loss rates are $(1,0.2)$ at the left boundary and $(0.2,1)$ at the right. These boundary conditions support a nonequilibrium profile and transport rather than independent bulk equilibration \mcite{prosen2011,znidaric2010,landi2022}.

For $N_s=5$, a trace-constrained sparse linear solve yields the reference: replace one row of the vectorized stationary equation by $\tr\rho=1$, solve the resulting linear system, and verify the answer with the original Liouvillian. At $N=6,\ldots,9$, the reference-generation routines propagate the exact full-generator semigroup using a sparse Krylov exponential, not the product-formula baseline. Consecutive propagation windows are of duration 8 for TFIM and 20 for Heisenberg. Their stopping criteria are $\norm{\Lcal\rho}_F<2\times10^{-12}$ and a change below $10^{-11}$ between successive windows. The output is Hermitized, trace normalized, and checked again.

At $N=10,11,12$, the TFIM references are computed by trace-constrained GMRES in the full orthonormal Pauli basis, with a diagonal preconditioner, relative tolerance $10^{-12}$, and absolute tolerance $10^{-14}$. The restart lengths are 40 at $N=10$ and 20 at $N=11,12$. The Heisenberg references at $N=10,11$ are computed by trace-constrained GMRES in all number sectors, with shifted incomplete-LU preconditioning. At $N=12$, an initialization in the exact reflection-plus-spin-flip invariant subspace is lifted to all number sectors and used in full-number-space, trace-constrained GPU GMRES, with absolute tolerance $10^{-12}$, zero relative tolerance, and restart length 20. The final solve returns a converged solver flag, and the reference is checked for trace, Hermiticity, positivity, and stationarity. The accepted large-size Heisenberg references satisfy $\norm{\Lcal\rho_{\rm ref}}_F^2<4\times10^{-24}$.

Reference states are cached and revalidated on loading. Table~\ref{tab:referencechecks} lists their checks through $N=12$. All traces agree with one to numerical precision, and all minimum eigenvalues are positive. The accompanying records specify the generating solver, representation conversions, and convergence checks.

\begin{table}[h]
\centering\small
\caption{Independent reference-state checks through $N=12$. The residual is squared. These checks concern numerical consistency, not an a priori residual-to-state-error theorem.}\label{tab:referencechecks}
\begin{tabular}{lrrrr}\toprule
Model & $N$ & $\norm{\Lcal\rho_{\rm ref}}_F^2$ & $\lambda_{\min}(\rho_{\rm ref})$ & $\tr\rho_{\rm ref}^2$\\\midrule
TFIM & 5 & $4.85\times10^{-30}$ & $1.09\times10^{-4}$ & 0.354935 \\
TFIM & 6 & $3.16\times10^{-24}$ & $1.51\times10^{-5}$ & 0.323819 \\
TFIM & 7 & $8.81\times10^{-28}$ & $2.08\times10^{-6}$ & 0.294870 \\
TFIM & 8 & $3.75\times10^{-24}$ & $2.90\times10^{-7}$ & 0.268980 \\
TFIM & 9 & $5.51\times10^{-29}$ & $4.04\times10^{-8}$ & 0.245835 \\
TFIM & 10 & $1.82\times10^{-29}$ & $5.66\times10^{-9}$ & 0.224982 \\
TFIM & 11 & $4.31\times10^{-29}$ & $7.91\times10^{-10}$ & 0.206070 \\
TFIM & 12 & $2.31\times10^{-28}$ & $1.10\times10^{-10}$ & 0.188846 \\
Heisenberg & 5 & $2.94\times10^{-31}$ & $5.80\times10^{-3}$ & 0.051202 \\
Heisenberg & 6 & $3.82\times10^{-24}$ & $1.60\times10^{-3}$ & 0.031516 \\
Heisenberg & 7 & $8.02\times10^{-28}$ & $4.76\times10^{-4}$ & 0.020026 \\
Heisenberg & 8 & $3.92\times10^{-27}$ & $1.25\times10^{-4}$ & 0.013107 \\
Heisenberg & 9 & $1.94\times10^{-26}$ & $3.51\times10^{-5}$ & 0.008801 \\
Heisenberg & 10 & $8.59\times10^{-26}$ & $8.90\times10^{-6}$ & 0.006036 \\
Heisenberg & 11 & $3.89\times10^{-26}$ & $2.40\times10^{-6}$ & 0.004209 \\
Heisenberg & 12 & $9.94\times10^{-25}$ & $5.94\times10^{-7}$ & 0.002970 \\

\bottomrule\end{tabular}
\end{table}

The Heisenberg calculation exploits the exact operator space block diagonal in total excitation number. All population sectors are retained; the reservoirs connect adjacent sectors. The space has dimension $\sum_{k=0}^N\binom Nk^2=\binom{2N}{N}$, rather than $4^N$. Individual blocks are \emph{not} normalized. For $\rho=\bigoplus_k\rho_k$ and $\sigma=\bigoplus_k\sigma_k$, squared fidelity is
\begin{equation}
 F(\rho,\sigma)=\left(\sum_k\tr\sqrt{\sqrt{\sigma_k}\rho_k\sqrt{\sigma_k}}\right)^2,
\end{equation}
not the sum of separately normalized sector fidelities \mcite{jozsa1994}. Full-Hilbert-space channel reconstructions of selected protocols agree with the block calculation to $4.5\times10^{-14}$ or better in Frobenius norm through $N=9$. Reference propagation, channel application, and optimization are separate computations.

\section{Physical channels, product-formula inclusion, and resources}
\label{sec:channels}
The three-group layer has chronological sequence $H_1,H_2,\Dcal$, i.e.,
\begin{equation}
 \Phi_\ell=e^{\tau_\ell\Dcal}e^{b_\ell\Hcal_2}e^{a_\ell\Hcal_1},
 \quad a_\ell,b_\ell\in[-\pi,\pi],\quad \tau_\ell\in[0,4].
\end{equation}
Each of $a_\ell,b_\ell,\tau_\ell$ multiplies an entire physical generator group, with its fixed site or bond coefficients. There are $3P$ parameters, not $3NP$. In Heisenberg, sharing the duration of the reservoir group preserves its left/right rate asymmetry. Spatial sharing is a control constraint, not an assertion that the prepared state is translation invariant. The coherent factors are exact unitary channels. In the TFIM, the $X$ rotations factorize by site and all $ZZ$ terms commute. In Heisenberg, the entire $XX+YY$ group is exponentiated in each number sector; at the largest sizes, an equivalent exact single-particle diagonalization and complex Givens factorization evaluates this same whole-group exponential. Neither implementation introduces a bondwise Trotter approximation. The sign of a coherent control is reversed when its parameter is negative. A hardware interpretation must therefore supply independent scheduling and the stated Hamiltonian-sign controls. Complete positivity follows from the construction, not from a constraint imposed after optimization \mcite{lindblad1976,gks1976,lloydviola2001}.

For a single-site pump/loss generator $u\Dcal[\sigma^+]+d\Dcal[\sigma^-]$, let $\eta=e^{-(u+d)\tau}$ and $p=u/(u+d)$. The exact linear channel on any local block is
\begin{align}
 \rho'_{00}&=[p+(1-p)\eta]\rho_{00}+p(1-\eta)\rho_{11},\\
 \rho'_{11}&=(1-p)(1-\eta)\rho_{00}+[1-p+p\eta]\rho_{11},\\
 \rho'_{01}&=\sqrt\eta\,\rho_{01},\quad \rho'_{10}=\sqrt\eta\,\rho_{10}.
\end{align}
These formulas also act on unnormalized environment-indexed blocks. For local amplitude damping, $u=0,\quad d=1$, a Kraus representation is $K_0=\ket1\bra1+\sqrt\eta\ket0\bra0$ and $K_1=\sqrt{1-\eta}\ket1\bra0$. Distinct-site dissipative channels commute. Pump and loss at the same site are included jointly through the exact channel above, not applied as two independently split factors.

\subsection{Explicit auxiliary-qubit dilation}
Channel dilation realizes an irreversible system map by a unitary on a larger system followed by a partial trace. Gate-based constructions for qubit channels and Markovian dynamics are given in Refs.~\mcite{wang2013channels,sweke2014,dive2015}. Here the local reservoir factors have a particularly simple exact implementation, so no small-exposure approximation is needed.

Use the ordered system--auxiliary basis $\ket{00},\ket{01},\ket{10},\ket{11}$. For pure loss, $\eta=e^{-d\tau}$, $c=\sqrt\eta$, and $s=\sqrt{1-\eta}$, one unitary extension is
\begin{equation}
 V_\eta=\begin{pmatrix}c&0&0&-s\\0&1&0&0\\0&0&1&0\\s&0&0&c\end{pmatrix},\quad
 \mathcal E_{\eta}^{\rm loss}(\rho)
 =\tr_a[V_\eta(\rho\otimes\ket0\bra0_a)V_\eta^\dagger].
 \label{eq:smdilation}
\end{equation}
Thus $K_j={}_a\bra jV_\eta\ket0_a$ yields precisely the Kraus operators above. For example, with $R_y(\beta)=e^{-\ii\beta Y/2}$ and $\alpha=\arccos\sqrt\eta$, a chronological gate sequence is: apply $X$ on the system, apply a system-controlled $R_y(2\alpha)$ on the auxiliary, apply $X$ on the system again, and apply a CNOT from auxiliary to system. This agrees with the required isometry on auxiliary input $\ket0_a$; its action on the unused input subspace can differ from the matrix in Eq.~\eqref{eq:smdilation}. A controlled rotation has a fixed decomposition into two CNOTs and one-qubit rotations, giving one explicit constant-depth construction with three CNOTs in total before further simplification. The auxiliary is then discarded or reset, with no postselection.

For simultaneous pump and loss, use $\eta=e^{-(u+d)\tau}$ and $p=u/(u+d)$ as above. The four exact Kraus operators are
\begin{align}
 G_0&=\sqrt{1-p}(\sqrt\eta\ket0\bra0+\ket1\bra1),\quad
 G_1=\sqrt{(1-p)(1-\eta)}\ket1\bra0,\nonumber\\
 G_2&=\sqrt p(\ket0\bra0+\sqrt\eta\ket1\bra1),\quad
 G_3=\sqrt{p(1-\eta)}\ket0\bra1.
 \label{eq:smgad}
\end{align}
They obey $\sum_\nu G_\nu^\dagger G_\nu=I$ and reproduce all four block formulas above. Equivalently, the channel is $(1-p)\mathcal E_\eta^{\rm loss}+p\mathcal E_\eta^{\rm pump}$ with $\mathcal E_\eta^{\rm pump}=\mathcal X\mathcal E_\eta^{\rm loss}\mathcal X$ and $\mathcal X(\rho)=X\rho X$. Sample a classical branch with probabilities $(1-p,p)$ and execute the corresponding one-auxiliary circuit; averaging unconditional outcomes realizes the desired density operator. Alternatively, a pure-environment dilation uses two auxiliaries and the isometry $\ket\psi\mapsto\sum_{\nu=0}^3G_\nu\ket\psi\otimes\ket\nu_a$, extended to a unitary. This is an exact implementation of the joint generator, not a sequential pump/loss splitting. At the left/right boundaries $p=5/6,1/6$, respectively, and $\eta=e^{-1.2\tau}$. If $u+d=0$ or $\tau=0$, the channel is the identity.

For TFIM, $\bigotimes_j\mathcal E_j$ can be applied in parallel with $N$ local auxiliaries, each reset before reuse. With one auxiliary, sequential coupling and reset uses fewer qubits but $O(N)$ dissipative-block depth, plus any routing cost. Heisenberg needs only the two boundary channels, which can operate in parallel with boundary auxiliaries. An auxiliary entangled with previous system outputs must not be reused without reset or replacement: retaining that memory would simulate a different channel.

\subsection{From grouped blocks to compiled depth}
For a common sequential template schedule, summing template depths gives the article's depth formulas; this is the depth of that explicit implementation before extra cross-template optimization. Removing zero factors can reduce the LVA count further. Equal block counts imply equal depth only when the template costs are equal. For the centered symmetric formula in the article, the depth difference is
\begin{equation}
 D_2-D_{\rm LVA}=(m+1-P)d_1+(2m-P)d_2+(m-P)d_D.
 \label{eq:smdepthdiff}
\end{equation}
For the $N=12$ sequential-transfer budgets, TFIM has $P=7$ against $m=9$, giving $3d_1+11d_2+2d_D>0$, and Heisenberg has $P=50$ against $m=96$, giving $47d_1+142d_2+46d_D>0$. Both inequalities hold before and after target refinement. An alternative Strang centering has its own template multiplicities and must be counted accordingly.

TFIM admits exact angle-controlled templates: all $X$ gates are parallel and the $ZZ$ gates occupy two disjoint-bond rounds on a linear device. With local auxiliary connectivity and parallel reset, all three template depths can be independent of $N$. The Heisenberg $XX+YY$ sum does not factor into commuting bond terms. Approximating it by even/odd product formulas can require angle- and accuracy-dependent subdivision; using a common fixed-depth template is an additional implementation assumption, not a property of the dense numerical exponential. If template depths vary, replace each product $n_ad_a$ by the actual sum over compiled occurrences, and apply the same accuracy and hardware constraints to both protocols.

Compilation errors also consume the endpoint-accuracy budget. If $\mathcal E_k$ is an ideal block and $\widehat{\mathcal E}_k$ its implemented channel, define $\delta_k=\frac12\norm{\widehat{\mathcal E}_k-\mathcal E_k}_\diamond$. A telescoping expansion and channel contractivity give, for the same input,
\begin{equation}
 \frac12\norm{\widehat\rho_{\rm out}-\rho_{\rm out}}_1
 \leq\sum_k\delta_k.
 \label{eq:smsynthesis}
\end{equation}
Therefore the ideal matched-fidelity budgets are not automatically hardware-certified budgets. A compiled comparison must check the final fidelity, allocating enough synthesis accuracy and including connectivity, reset, and noise. The explicit depth model supports a hardware advantage from fewer executions of the same templates; it does not turn the present noiseless data into a measured gate-depth benchmark.

A first-order digital step is obtained by setting $(a,b,\tau)=(h,h,h)$. A symmetric step with chronological sequence $H_1(h/2),H_2(h/2),\Dcal(h),H_2(h/2),H_1(h/2)$ is embedded through
\begin{equation}
 \begin{pmatrix}a_1&b_1&\tau_1\\a_2&b_2&\tau_2\\a_3&b_3&\tau_3\end{pmatrix}
 =\begin{pmatrix}h/2&h/2&h\\0&h/2&0\\h/2&0&0\end{pmatrix}.
 \label{eq:embedding}
\end{equation}
The equality holds at the channel level, not just in a small-$h$ expansion. It is verified numerically for both models. If a required duration lies outside the prescribed box, it must be subdivided into more admissible blocks. Product-formula inclusion is therefore an expressivity statement with sufficient depth and matching controls, not equal nominal-depth dominance. This distinction also applies to physically structured Hamiltonian ans\"atze \mcite{wecker2015,wiersema2020}.

We count $B=3P$ nominal factors for an LVA. A first-order sequence of $m$ steps costs $3m$. A symmetric second-order sequence costs $5m-(m-1)=4m+1$ after adjacent equal half-steps are merged. For LVA the main figures retain identities conservatively; a separate operation ledger removes factors below $10^{-9}$ and merges adjacent factors of the same generator. No cross-generator commutation or unexplained circuit simplification is assumed.

Neither count measures wall-clock time. At fixed sequential control amplitudes and no overhead, the LVA duration would be $T_{\rm seq}=\sum_\ell(|a_\ell|+|b_\ell|+\tau_\ell)$, whereas a three-group product formula has total sequential duration $3T$. A native simultaneous Lindblad evolution instead has duration $T$. Compilation, parallel scheduling, and reset overhead can change all these comparisons. Algorithms for approximating dynamical channels solve a different resource problem from endpoint control \mcite{kliesch2011,childsli2017,clevewang2017}.

\section{Classical source training and continuation}
\label{sec:source}
\subsection{Source-infidelity training}
The five-spin source protocols in Fig.~2 minimize $1-F$, with $F$ the squared Uhlmann fidelity. Every L-BFGS-B call has an iteration cap of 1000, including augmented candidates, plain INTERP stages, and the five random starts at each sampled depth. The bounds are $a,b\in[-\pi,\pi]$ and $\tau\in[0,4]$, with \texttt{gtol}$=10^{-10}$, \texttt{ftol}$=10^{-14}$, and \texttt{maxls}$=30$. Runs start at depth one and reach $P=20$ for TFIM and $P=50$ for Heisenberg. Source residual and Strang schedules supply additional initial guesses, which are optimized using infidelity. The unchanged identity-insertion candidate remains eligible in both models. Solver termination before 1000 iterations is allowed; reaching the cap is not treated as convergence.

The lower panels use the same augmented-continuation infidelities as the upper panels. All augmented, plain, and random endpoints are independently propagated before plotting, with unit trace and positivity checked to numerical precision. Full parameters, optimizer logs, validation records, and plotted coordinates are provided in the accompanying data and code. Main Figs.~3--6 use the same infidelity-trained source pool. The candidate rules and effort table below describe these source-training runs.

\subsection{Objectives and exact adjoint gradients}
For source training, set $\sigma=\rho_{\rm ref}^{(5)}$, $X=\sqrt\sigma\rho_M\sqrt\sigma$, and $f=\tr\sqrt X$. The loss is $\mathcal C_s=1-f^2$. At full-rank $X$, its terminal adjoint is $A_M=-f\sqrt\sigma X^{-1/2}\sqrt\sigma$. The implementation uses the spectral pseudoinverse of $X^{1/2}$ on eigenvalues larger than $\max(10^{-16},10^{-12}\lambda_{\max}(X))$ for rank-deficient shallow-circuit states; the objective itself remains the unregularized infidelity. Gradient checks at $P=1,3,7$ for both models give a maximum finite-difference gradient discrepancy of $2.42\times10^{-6}$ and a maximum loss-versus-independent-fidelity discrepancy of $1.01\times10^{-9}$.

The source reference does not enter any large-system local-cost optimization. Write $\rho_k=\mathcal E_k(\theta_k)\rho_{k-1}$ with $\mathcal E_k=e^{\theta_k\mathcal G_k}$ and propagate backward by
\begin{equation}
 A_{k-1}=\mathcal E_k^\dagger(A_k),\quad
 \partial_{\theta_k}\mathcal C=\operatorname{Re}\tr[A_k^\dagger\mathcal G_k(\rho_k)].
 \label{eq:adjoint}
\end{equation}
For the raw residual $\mathcal C_{\rm res}=\norm{\Lcal(\rho_M)}_F^2$, the terminal adjoint is $2\Lcal^\dagger\Lcal(\rho_M)$. For the local objective, with $T_A(\rho)=\tr_{\bar A}\rho$, it is
\begin{equation}
 A_M=\frac{2}{N-r+1}\Lcal^\dagger
     \sum_A T_A^\dagger T_A\Lcal(\rho_M),\quad
 T_A^\dagger(X_A)=X_A\otimes I_{\bar A},
 \label{eq:localadjoint}
\end{equation}
with the appropriate tensor ordering. There is no dimension-normalization factor in $T_A^\dagger$. Exact dense and sector implementations, together with finite differences, test both formulas. The numerical optimizer is L-BFGS-B \mcite{byrd1995}; the use of analytic gradients here is a classical implementation choice.

\subsection{Candidate construction and source-only selection}
A source continuation begins at $P=1$. The deterministic starting vector is $(0.1,0.1,0.1)$; an additional seeded starting vector samples coherent controls uniformly from $[-1,1]$ and the dissipative duration from $[0.1,1]$. At each later depth, the INTERP rule is applied separately to $a$, $b$, and $\tau$ \mcite{zhou2020,mele2022}. The previous optimized schedule also supplies an appended zero layer. This identity insertion exactly preserves the state before subsequent reoptimization.

At depths $3,5,7,9,12,16,20$ we additionally perturb the INTERP initialization with independent Gaussian deviations of standard deviations $0.07$ for coherent parameters and $0.025$ for dissipative durations, clipping durations to $[0,4]$. The Heisenberg extension adds perturbations at $P=24,30$. A precomputed source residual-INTERP sequence and, when available, an embedded source Strang solution provide further candidate starts. Residual pretraining is charged separately below. These are all source-system calculations; no target error is consulted when constructing or selecting a source candidate.

Each candidate is optimized for at most 1000 iterations, with \texttt{gtol}$=10^{-10}$, \texttt{ftol}$=10^{-14}$, and at most 30 line-search steps. The final candidate is selected by source infidelity. Both implementations explicitly retain the unchanged identity-insertion candidate alongside the optimized candidates, so the selected source infidelity is nonincreasing with depth. Maximum-iteration termination is retained as such, not relabeled as convergence. The primary source schedules have depths $P\leq20$ for TFIM and $P\leq50$ for Heisenberg.

Plain INTERP starts from $(0.1,0.1,0.1)$ and uses only interpolation followed by the same 1000-iteration optimization. Five independent random starts are also trained at $P=3,5,7,10,15,20$, with additional Heisenberg depths $30,40,50$. They sample coherent parameters uniformly from $[-\pi,\pi]$ and durations from $[0,4/P]$. The source-random comparison retains the lowest-source-loss candidate at each tested depth. All source selection precedes target evaluation; no target fidelity is used to choose a source schedule.

\subsection{Training effort}
The additional candidate searches consume real classical work. Table~\ref{tab:effort} reports objective/gradient evaluations, not just outer iterations. An evaluation returns the loss and its analytic gradient together. The augmented totals include every candidate optimized along the depth chain. Residual pretraining, the reference solve, and the source digital time search are distinct costs; the last two are not included in the evaluation columns. Plain continuation and direct random runs therefore are not equal-total-budget competitors.

\begin{table}[h]
\centering\small
\caption{Recorded classical source evaluations for the infidelity-trained protocols (1000-iteration cap per optimizer call). ``Residual seed'' counts the residual pretraining chain, with a 100-iteration cap per call, through the same maximum depth. ``Random'' sums five starts at each tested depth, not five at every intermediate depth.}\label{tab:effort}
\begin{tabular}{lrrrrrr}\toprule
 & $P_{\max}$ & Candidates & Augmented & Residual seed & Plain INTERP & Random\\\midrule
TFIM & 20 & 85 & 58031 & 2059 & 13779 & 18288 \\
Heisenberg & 50 & 189 & 159749 & 5099 & 43537 & 30950 \\

\bottomrule\end{tabular}
\end{table}

The source-state metric does not remove all conditioning issues. Its derivative follows the channel Jacobian, and dissipation in a long suffix can suppress sensitivity to early parameters. Moreover, a good source-state solution can transfer less well than another source solution. The reported selection rule is nonetheless fixed at the source; it is not retrospectively chosen by a large-system fidelity. Existing warm-start and HVA trainability results have their own assumptions \mcite{grant2019,skolik2021,mele2022,park2024,puig2025}. The response analysis below treats these finite-angle protocols directly, separately from the optimizer traces.

The five independent infidelity-trained random starts in main Fig.~2 are displayed as a range with their best source infidelity; neither a high-quality outlier nor a poor outlier is suppressed. More extensive statistics would be needed for a claim of asymptotic reliability or a matched-compute learning advantage.

\section{Reference-free target objectives and measurement}
\label{sec:target}
\subsection{Locality of the stationarity constraints}
For a finite-range Lindbladian $\Lcal=\sum_X\ell_X$ and a region $A$, let $R_A=\tr_{\bar A}\Lcal(\rho)$. In a Pauli basis on $r$ sites,
\begin{equation}
 R_A=2^{-r}\sum_{Q\in\mathcal P(A)}q_Q Q,\quad
 q_Q=\tr\{\rho\Lcal^\dagger(Q\otimes I_{\bar A})\}.
\end{equation}
The coefficients are real because the adjoint generator preserves Hermiticity. The identity coefficient vanishes by trace preservation. Orthogonality immediately yields
\begin{equation}
 \norm{R_A}_F^2=2^{-r}\sum_Qq_Q^2.
 \label{eq:pauliderivation}
\end{equation}
A term $\ell_X^\dagger(Q\otimes I)$ vanishes when $X$ is disjoint from $A$, since a trace-preserving local generator has $\ell_X^\dagger(I_X)=0$. Thus only terms touching $A$ survive, and their support lies in the union of $A$ and the interaction neighborhoods that touch it. For nearest-neighbor generators this involves at most the region and its immediately adjacent sites. The number of potentially different local constraints scales as $O(N4^r)$ at fixed interaction range, with coefficients determined by the physical generator. This is an observable-support statement, not a bound independent of coupling magnitudes or measurement variance \mcite{poulin2010,barthel2012}.

Local Pauli grouping, randomized measurement, or classical shadows can estimate the required expectations \mcite{huang2020,elben2023}. Unlike global purity and overlap estimates, which may use multi-copy or randomized constructions \mcite{elben2018,brydges2019}, the elementary expectations here are single-copy observables. The nonlinear loss still requires repeated preparations. Overlapping regions and shared measurement settings can reduce redundant measurements, but the resulting covariances must be included in a variance analysis.

\subsection{Avoiding the bias of squared sample means}
Suppose $X$ is an unbiased scalar estimator of $q$ with variance $v$. A naive sample mean $\bar X$ from $K$ preparations gives $\mathbb E\bar X^2=q^2+v/K$. Two independent batches instead give the unbiased product $\bar X_A\bar X_B$, with
\begin{equation}
 \mathbb E(\bar X_A\bar X_B)=q^2,\quad
 \operatorname{Var}(\bar X_A\bar X_B)=\frac{2q^2v}{K}+\frac{v^2}{K^2}.
 \label{eq:batches}
\end{equation}
An alternative using one batch of $K\geq2$ independent samples is the second-order U-statistic
\begin{equation}
 \widehat{q^2}=\frac{(\sum_{a=1}^KX_a)^2-\sum_{a=1}^KX_a^2}{K(K-1)}.
\end{equation}
It can be negative on an individual finite sample even though its expectation is nonnegative. Summing these estimators with the weights in Eq.~\eqref{eq:pauliderivation} gives an unbiased local-cost estimator. Treating the squares as noiseless during optimization would hide a systematic offset, particularly near stationarity.

The gradient satisfies $\partial_\theta\mathcal C_{\rm loc}^{(r)}=2[(N-r+1)2^r]^{-1}\sum q_Q\partial_\theta q_Q$. Independent estimators of the two factors avoid an analogous product bias. Unitary parameter-shift rules, generalized shift constructions, or stochastic gradient methods can be used when their assumptions fit the implemented channel \mcite{schuld2019,wierichs2022,banchi2021,spall1992}. The reported target refinements use exact expectations and the adjoint gradient in Eq.~\eqref{eq:localadjoint}.

\subsection{Transfer paths and largest-device training cost}
Main Fig.~5 compares direct and sequential transfer of infidelity-trained source protocols, with source and target iteration caps of 1000. Both paths use the $N_s=5$ source controls: TFIM at $P=7$ and Heisenberg at $P=20$. Target refinement minimizes the local Lindblad residual with $r=2$ and $r=4$, respectively, using the parameter bounds and stopping criteria in Sec.~\ref{sec:source}. Final optimizer parameters are retained regardless of their diagnostic target infidelity. All 28 target records, their initial and final parameters, objective histories, reference checks, and plotted points are provided in the accompanying data and code.

For direct transfer, every target $N=6,\ldots,12$ is initialized with the same five-spin source vector. For unit-step sequential transfer, the first target $N=6$ has the identical initialization, and every subsequent target $N$ inherits the complete final parameter vector from the local refinement at $N-1$. The ``before'' value is evaluated before local optimization at the displayed size. In particular, the sequential before-refinement state at $N=12$ is the frozen transfer of the controls obtained at $N=11$ along the sequential path. Those controls inherit the refinements at sizes 6 through 11 and are held fixed when applied to the 12-spin system; no iteration, local-objective evaluation, or parameter update from the 12-spin refinement is used. The subsequent 12-spin refinement is shown separately by the after-refinement curve.

The two target-12 preparations compared in Table~\ref{tab:largestdevice} use equal source depth and the same source-training pool. The sequential before-refinement infidelity is slightly lower for TFIM and lower for Heisenberg than direct transfer followed by 1000 iterations at $N=12$. Infidelity is used to validate this comparison, not as a target-training loss or an intermediate-iterate selection rule.

\begin{table}[ht]
\centering\small
\caption{Target-$12$ comparison and optimizer effort. $Q$ counts recorded objective-function evaluations, not quantum shots. Sequential preparation stops training at $N=11$; hence its $Q_{12}=0$ in this comparison. The shared classical source training is excluded from both columns.}\label{tab:largestdevice}
\begin{tabular}{lrrrrr}\toprule
Model & $P$ & Sequential before $1-F$ & Direct after $1-F$ & Sequential $Q_{N<12}$ & Direct $Q_{12}$\\\midrule
TFIM & 7 & 0.009119968 & 0.009308256 & 6425 & 1090\\
Heisenberg & 20 & 0.186136602 & 0.224825156 & 6573 & 1092\\\bottomrule
\end{tabular}
\end{table}

The sequential route uses 5873 intermediate-size iterations for TFIM and 6000 for Heisenberg, compared with 1000 iterations on the 12-spin device for direct target refinement. It therefore need not lower the total training cost. It can reduce training cost on the largest quantum hardware by replacing its variational-training calls with optimization on smaller instances. A hardware accounting would weight each size's evaluation count by the associated circuit, measurement, gradient, and shot costs; the present evaluation counters do not constitute measured hardware runtimes.

LVA does not require a unit-step path. The same rule can be applied to any chosen increasing sequence of system sizes, with refinement performed only at its selected intermediate points and optionally at the final size. Table~\ref{tab:coarseroutes} compares the $N=12$ endpoints for direct transfer, $5\to12$, and the two coarse routes $5\to10\to12$ and $5\to6\to8\to10\to12$. At every selected target size, local-residual refinement starts from the final controls of the preceding size in that route. The source protocols, local objectives, control bounds, and stopping tolerances are the same as in the direct and unit-step tests. Each target refinement has a cumulative iteration budget of at most 1000. Full route records and numerical validation are provided in the accompanying data and code.

\begin{table}[ht]
\centering\small
\caption{Infidelity at $N=12$ for direct transfer and two coarse transfer routes. Before refinement uses the five-spin source controls for direct transfer and the controls optimized at $N=10$ along the corresponding coarse route; after refinement uses the final local-residual optimization parameters at $N=12$. TFIM uses $P=7$, and Heisenberg uses $P=20$. Source training and target refinement have iteration budgets of 1000. These are finite-budget endpoints, not certified converged optima.}\label{tab:coarseroutes}
\begin{tabular}{llrr}\toprule
Model & Route & $1-F$ before refinement & $1-F$ after refinement\\\midrule
TFIM & $5\to12$ & 0.022360784 & 0.009308256\\
TFIM & $5\to10\to12$ & 0.009124529 & 0.009309287\\
TFIM & $5\to6\to8\to10\to12$ & 0.009121710 & 0.009303775\\
Heisenberg & $5\to12$ & 0.457970805 & 0.224825156\\
Heisenberg & $5\to10\to12$ & 0.235510256 & 0.204853852\\
Heisenberg & $5\to6\to8\to10\to12$ & 0.215470545 & 0.213169911\\
\bottomrule\end{tabular}
\end{table}

For TFIM, both coarse routes improve the infidelity before refinement at $N=12$: direct transfer gives 0.022360784, compared with 0.009124529 for $5\to10\to12$ and 0.009121710 for $5\to6\to8\to10\to12$. Adding the intermediate sizes $N=6,8$ to $5\to10\to12$ also lowers the infidelity after refinement at $N=12$, from 0.009309287 to 0.009303775; the longer route improves on the directly refined value of 0.009308256 as well. For Heisenberg, with the same 1000-iteration refinement budget at $N=12$, both coarse routes improve on the direct-transfer endpoint (0.224825), reaching 0.204854 and 0.213170, respectively, with the shorter route giving the lower value. These tests show that refinement at intermediate sizes can improve endpoint accuracy, and adding further intermediate sizes can provide additional gains, as observed in TFIM. The magnitude of the improvement depends on the model and the chosen route.

\subsection{Local accuracy and measurement resources}
Reducing the local stationarity cost need not improve global fidelity. For example, direct TFIM transfer at $N=6,P=7$ changes the infidelity from $0.002809$ to $0.003714$ after two-site refinement, despite lowering the local cost. Increasing the region size adds correlation constraints, but also changes the finite-budget optimization landscape and measurement variance; it is not a universal guarantee of improved global accuracy.

One may supplement the local stationarity objective with magnetization profiles, currents, or a larger-region hierarchy, provided the quantities used for training and those used only for validation are clearly separated. Agreement of a few observables does not replace a global fidelity calculation when that metric is reported. Short continuation under the target Lindbladian supplies a drift diagnostic, but a metastable state can also drift slowly \mcite{macieszczak2016}. The numerical tests therefore assess the local objective through reference-state diagnostics at accessible sizes, rather than claiming an experimentally available global certificate.

The observable support in the cost remains bounded as $N$ increases at fixed $r$. Nevertheless, shot requirements, the number of parameters, and the number of repeated optimizations are separate resources. Deep dissipative suffixes can erase parameter sensitivity, while uncontrolled device noise can further suppress gradients \mcite{mcclean2018,cerezo2021cost,wang2021}. The designed dissipation in LVA must not be identified with that hardware noise. Noise-aware training is compatible with the framework but is not simulated here.

\section{Digital-relaxation baselines and matched-accuracy budgets}
\label{sec:baseline}
\subsection{Sequential-transfer budgets through twelve spins}
Main Fig.~6 uses source-infidelity training and sequential local-residual refinement, with 1000-iteration caps per source and target optimizer call. The residual regions are $r=2$ for TFIM and $r=4$ for Heisenberg. At each depth, the before-refinement state at size $N>6$ uses the controls refined at the preceding size; it is not direct frozen transfer from the five-spin source. Additional source depths used only for the resource sweeps continue the saved five-spin protocols by repeated INTERP or exact zero-layer insertion, with a source-only Strang-informed candidate when available; all optimized candidates still minimize infidelity with a 1000-iteration cap. Only completed optimizer outputs enter the resource comparison. Target fidelity is used for assessment, not for selecting an intermediate target iterate.

The source and target records, plotted coordinates, and digital-baseline time scans are provided in the accompanying data and code. Table~\ref{tab:sequentialbudgets} reports the smallest successful sampled block budgets, using all completed schedules. The source points at $N=5$ have no target refinement. Missing intermediate depths are not inferred to fail, and these sampled minima are not certified global optima.

\begin{table}[t]
\centering\small
\caption{Sequential-transfer sampled budgets in main Fig.~6. Before and after refer to refinement at the displayed size. All counts are nominal grouped blocks; a dash denotes the source point without target refinement.}
\label{tab:sequentialbudgets}
\begin{tabular}{lrrrrrr}\toprule
 & & & \multicolumn{2}{c}{Before refinement} & \multicolumn{2}{c}{After refinement}\\
Model & $N$ & $\widehat B_2$ & $\widehat B_{\rm LVA}$ & $\widehat G_B^{(2)}$ & $\widehat B_{\rm LVA}$ & $\widehat G_B^{(2)}$\\\midrule
TFIM & 5 & 25 & 15 & 1.667 & -- & -- \\
TFIM & 6 & 29 & 18 & 1.611 & 18 & 1.611 \\
TFIM & 7 & 29 & 18 & 1.611 & 18 & 1.611 \\
TFIM & 8 & 33 & 18 & 1.833 & 18 & 1.833 \\
TFIM & 9 & 33 & 21 & 1.571 & 21 & 1.571 \\
TFIM & 10 & 33 & 21 & 1.571 & 21 & 1.571 \\
TFIM & 11 & 37 & 21 & 1.762 & 21 & 1.762 \\
TFIM & 12 & 37 & 21 & 1.762 & 21 & 1.762 \\
Heisenberg & 5 & 41 & 18 & 2.278 & -- & -- \\
Heisenberg & 6 & 61 & 33 & 1.848 & 30 & 2.033 \\
Heisenberg & 7 & 89 & 45 & 1.978 & 39 & 2.282 \\
Heisenberg & 8 & 121 & 60 & 2.017 & 60 & 2.017 \\
Heisenberg & 9 & 161 & 66 & 2.439 & 66 & 2.439 \\
Heisenberg & 10 & 217 & 108 & 2.009 & 90 & 2.411 \\
Heisenberg & 11 & 289 & 108 & 2.676 & 108 & 2.676 \\
Heisenberg & 12 & 385 & 150 & 2.567 & 150 & 2.567 \\
\bottomrule\end{tabular}
\end{table}

For legibility, the two $N=12$ resource panels display a subset of the evaluated points. The plotted infidelity ranges are $[10^{-5},1.2]$ for TFIM and $[0.02,1.2]$ for Heisenberg. All completed schedules and evaluated baseline samples, including points outside the displayed ranges, enter the threshold-budget and gain calculations.

At $N=12$, the successful TFIM budgets are $B=21$ at both stages against $B_2=37$, with infidelities $0.00911997$ before and $0.00931177$ after refinement. Heisenberg uses $B=150$ at both stages against $B_2=385$, with infidelities $0.03889694$ and $0.03161480$. The corresponding symmetric second-order crossings are $m=9$ and $m=96$, respectively, in the centering used in the article. These are finite-grid, time-optimized baseline crossings. The baselines use grouped-channel propagation and an independent search over total evolution time. The usual time-search ranges are $[0,40]$ for TFIM and $[0,60]$ for Heisenberg, with grid spacing one and an additional spacing-$0.2$ grid on $[0,2]$. Competitive local minima are refined by bounded scalar minimization, examining up to six brackets. Some additional $N=12$ Heisenberg baseline sweeps extend to $T=120$, using spacing two from 4 to 60 and additional points at 80, 100, and 120. The saved time samples, centering choice, and generating file are retained for each record; the aggregate keeps the lowest recorded infidelity at each order and nominal block budget.

\subsection{Time optimization and comparison scope}
For every tested repetition count $m$, the baseline starts from the same all-down state and minimizes the global infidelity over its total evolution time $T$. The first-order sequence is $(e^{h\Dcal}e^{h\Hcal_2}e^{h\Hcal_1})^m$, $h=T/m$. The symmetric sequence is the five-factor formula specified in Sec.~\ref{sec:channels}; its reversed centering is also evaluated around the threshold. This implements the standard first-/second-order distinction without attributing a finite-step fixed point to the target generator \mcite{trotter1959,strang1968,suzuki1976}.

Time grids have spacing one, augmented by a grid of spacing $0.2$ on $[0,2]$. The main search ranges are $[0,40]$ for TFIM and $[0,60]$ for Heisenberg. Local minima within 20\% of the best grid value, with an absolute $10^{-11}$ allowance, are refined by bounded scalar minimization. Up to six such brackets are examined, with time tolerance $2\times10^{-4}$; threshold checks use $10^{-4}$. Every sampled time and its infidelity is retained. Far from a crossing the plotted curve may use a sparser $m$ grid. Neither bounded time scans nor a finite depth grid certify a global optimum.

A more advanced Lindbladian simulation method would be a different comparator \mcite{kliesch2011,childsli2017,clevewang2017}; the present results do not establish dominance over all simulation algorithms.

The finite target-refinement depth grid means that its threshold budget is a sampled value. Missing an intermediate depth cannot be interpreted as a proof that that depth fails. Conversely, a displayed success is an explicitly evaluated state, not interpolation between fidelity values. The use of the target reference to optimize baseline time makes the comparison favorable to the relaxation baseline; LVA target parameters are optimized using only local stationarity.

The block gains do not include source pretraining, reference construction, target shots, or compilation. They are relevant when preparation circuits are reused, but a full amortized comparison must specify how many target states are prepared and how many target refinements are required. A fixed number of grouped controls also need not mean a size-independent physical circuit depth. These distinctions prevent a finite-size state-preparation comparison from being interpreted as a universal quantum computational speedup \mcite{preskill2018,cerezo2021review}.

\section{Modal diagnostics and preparation constraints}
\label{sec:mechanism}

\subsection{Accessible operator spaces and decay-rate groups}
Main Fig.~3 uses $N=5$ throughout and the same source-infidelity training pool as main Fig.~2, with a 1000-iteration cap per optimizer call. The TFIM and boundary-driven Heisenberg schedules have $P=7$ and $P=20$, respectively. Natural-relaxation trajectories are computed with the exact full generator; LVA outputs are recorded after each individual generator block, including zero-duration factors. The source controls, modal coefficients, trajectories, and plotted coordinates are provided in the accompanying data and code. Both local-chain LVA series are directly propagated from saved controls.

At $N=5$, reflection-invariant operators give dimension 544 for local TFIM, and operators block diagonal in excitation number give dimension 252 for Heisenberg. These are exact accessible operator spaces for the stated generators and input, not product-state or fixed-filling approximations. The generator leakage outside each chosen space is numerically negligible.

We use right eigenoperators with unit Frobenius norm and dual left eigenoperators. The stationary mode is removed before forming the group norms $W_g$. With $\Gamma=-\operatorname{Re}\lambda$ and $\Gamma_s$ the slowest accessible nonzero rate, the four groups are $\Gamma=\Gamma_s$, $\Gamma_s<\Gamma<a$, $a\leq\Gamma<b$, and $b\leq\Gamma$. Table~\ref{tab:modalgroups} specifies the cutoffs. Numerically coincident slow rates are collected within a tolerance of order $10^{-6}$ in the stated rate units; other bin boundaries use a $10^{-7}$ tolerance. Degenerate modes remain in the same group. The weights refer to the fixed numerical eigenbasis: unit normalization does not make them invariant under arbitrary nonunitary rebasings of a degenerate space. They are used as within-model trajectory diagnostics, not as comparable error probabilities across models.

\begin{table}[ht]
\centering\small
\caption{Decay-rate grouping for main Fig.~3. Rates are in the reference units of each model; all reference rates are set to one. Counts exclude the stationary mode.}\label{tab:modalgroups}
\begin{tabular}{lrrrrl}\toprule
Model & Operator dimension & $\Gamma_s$ & $a$ & $b$ & Group counts\\\midrule
Local TFIM & 544 & 0.513113 & 1.5 & 2.5 & $(1,15,243,284)$\\
Boundary-driven Heisenberg & 252 & 0.245877 & 0.8 & 1.5 & $(1,24,186,40)$\\\bottomrule
\end{tabular}
\end{table}

The right-eigenvector condition numbers are approximately 147 and 15.1, respectively. Relative eigen-equation residuals are below $10^{-14}$. Reconstruction of the propagated local-chain deviations from their modal coefficients has maximum relative Frobenius error $5.6\times10^{-14}$ for TFIM and $1.1\times10^{-14}$ for Heisenberg. At the endpoint, the four group weights are approximately $(2.76\times10^{-3},1.81\times10^{-2},6.52\times10^{-2},2.45\times10^{-2})$ for TFIM and $(7.63\times10^{-5},3.78\times10^{-3},6.80\times10^{-3},7.79\times10^{-4})$ for Heisenberg. The directly propagated physical states reproduce the source infidelities $0.00169981828$ and $0.000264253310$. The display floor $10^{-4}$ censors small values only in the graphic; the data retain their full numerical values.

\subsection{Mode conventions and the exact affine decomposition}
For a diagonalizable target generator, choose dual modes with $\tr(Q_k^\dagger R_j)=\delta_{kj}$ and $c_k=\tr[Q_k^\dagger(\rho-\rho_{\rm ss})]$. Natural evolution gives $c_k(t)=e^{\lambda_k t}c_k(0)$; an initial rotation changes the coefficients, not the rates \mcite{carollo2021,kochsiek2022}. Write $U_\ell=e^{-\ii b_\ell H_2}e^{-\ii a_\ell H_1}$, $\mathcal U_\ell(X)=U_\ell XU_\ell^\dagger$, $\mathcal E_\ell=e^{\tau_\ell\Dcal}$, and $\Phi_\ell=\mathcal E_\ell\mathcal U_\ell$. With $X_{\ell-1}=\rho_{\ell-1}-\rho_{\rm ss}$, linearity gives
\begin{align}
 X_\ell&=A_\ell+\Delta_\ell,\quad
 A_\ell=\Phi_\ell(X_{\ell-1}),\quad
 \Delta_\ell=\Phi_\ell(\rho_{\rm ss})-\rho_{\rm ss},\label{eq:affineerror}\\
 c_{\ell k}&=\sum_{j\ne0}M_{\ell,kj}c_{\ell-1,j}+d_{\ell k},\quad
 M_{\ell,kj}=\tr[Q_k^\dagger\Phi_\ell(R_j)],\quad
 d_{\ell k}=\tr(Q_k^\dagger\Delta_\ell).
\end{align}
The generally signed or complex coefficients are not probabilities. For nonorthogonal modes, $\norm{X}_F^2=\mathbf c^\dagger G\mathbf c$ with $G_{kj}=\tr(R_k^\dagger R_j)$, rather than $\sum_k|c_k|^2$. The stationary coefficient is one for the state and zero for its deviation $X$. Target-$\Lcal$ eigenmodes also need not be contraction directions of $\Dcal$.

Equivalently, $\Phi_\ell=\mathcal U_\ell\widetilde{\mathcal E}_\ell$ with $\widetilde{\mathcal E}_\ell=\mathcal U_\ell^{-1}\mathcal E_\ell\mathcal U_\ell$ has rotated jumps $U_\ell^\dagger L_\mu U_\ell$. A cumulative frame $W_\ell=U_\ell\cdots U_1$ conjugates each dissipative channel by $W_\ell$. Neither identity changes the target generator or requires a Lindblad logarithm of the layer.

\subsection{Dissipative spectral preimages and terminal-time diagnostics}
The main-text preimage diagnostic uses the infidelity-trained TFIM protocol at $N=5,P=7$, with a 1000-iteration optimization cap. Its curves are evaluated directly from the learned controls. Throughout this analysis $\rho_{\rm ss}$ denotes the target, $\rho_{\rm pre}$ the physical input to the final dissipative block, and $\mathcal P_g$ an algebraic spectral projector of $\Dcal$. These projectors differ from the full-$\Lcal$ mode groups used for main Fig.~3 and from the Pauli-string set $\mathcal P(A)$ in Sec.~\ref{sec:target}.

For $\Dcal\mathcal P_g=-\Gamma_g\mathcal P_g$, define $C_g=\mathcal P_g\rho_{\rm ss}$ and $A_g=\mathcal P_g\rho_{\rm pre}$. Applying $\mathcal P_g$ to $(\Hcal+\Dcal)\rho_{\rm ss}=0$ gives $\Gamma_gC_g=\mathcal P_g\Hcal(\rho_{\rm ss})$. A positive-rate target component is therefore compatible with stationarity of the full generator. For $\Gamma_g>0$ and nonzero $A_g,C_g$, define the relative mismatch $r_g(t)=\norm{e^{-\Gamma_gt}A_g-C_g}_F/\norm{C_g}_F$ and introduce
\begin{equation}
 x_g(t)=e^{-\Gamma_gt}\frac{\norm{A_g}_F}{\norm{C_g}_F},\quad
 \cos\vartheta_g=\frac{\operatorname{Re}\tr(A_g^\dagger C_g)}{\norm{A_g}_F\norm{C_g}_F}.
 \label{eq:smpreimageangle}
\end{equation}
Expansion of the Frobenius norm gives
\begin{equation}
 r_g^2(t)=[x_g(t)-1]^2+2x_g(t)(1-\cos\vartheta_g).
 \label{eq:smpreimageerror}
\end{equation}
Thus both a compatible decay time and small directional mismatch are needed. With $q_g=\operatorname{Re}\tr(A_g^\dagger C_g)/\norm{A_g}_F^2$, the interior optimum for $0<q_g<1$ is
\begin{equation}
 t_g^*=-\Gamma_g^{-1}\ln q_g.
 \label{eq:smpreimagetime}
\end{equation}
For $q_g\geq1$ the optimum on $t\geq0$ is at zero; for $q_g\leq0$ the infimum is approached as $t\to\infty$. A zero $A_g$ gives a time-independent mismatch. Zero-rate groups retain $A_g-C_g$ throughout the final block. Since algebraic spectral projectors need not be orthogonal in the Hilbert--Schmidt inner product, their squared errors do not generally add to the full state error or infidelity. This simple exponential description assumes a diagonalizable dissipator with real rates; noncommuting dissipative subblocks require analysis of their full finite-time channel.

The calculation independently constructs the five-spin TFIM Hamiltonian and sparse Liouvillian, solves the trace-constrained stationary equation, and replays every coherent and dissipative block. Site-local amplitude-damping projectors have rates $0,1/2,1$; their tensor products yield all eleven global groups $\Gamma_g=0,1/2,\ldots,5$. All groups are retained in reconstruction. The three displayed rates are selected from the target spectrum, independently of the mismatch curves or their minima. We retain positive-rate groups with $\norm{C_g}_F$ at least 10\% of the largest positive-rate target norm, then choose the slowest and fastest retained rates and the rate closest to their midpoint. A midpoint tie is resolved by the larger target norm. This gives $\Gamma_g=0.5,2,4$; the weak-target $4.5$ and $5$ groups are excluded from display but retained in the calculation. The selected target norms are $0.7416635,1.2139943,0.2081261$. Their normalized pre-dissipation amplitudes are $1.113724,1.572232,2.512367$, while the required amplitudes are $1.121818,1.583761,2.508300$. The independent optimal times divided by $\tau_P=0.2299013005$ are $0.938374,0.984839,1.002717$. The norm threshold only selects representative nonnegligible target components; it is not a probability cutoff.

The scan fixes every earlier parameter and $\rho_{\rm pre}$, evaluating $t/\tau_P$ from 0 to 2 in steps of 0.005. The comparison input is obtained from exact $e^{s\Lcal}\rho_0$ at $s=1.6291005$, selected solely to match the pre-dissipation infidelity $0.038008997$. The same $e^{t\Dcal}$ is then applied to each input. At $t=\tau_P$ their endpoint infidelities are $0.00169981828$ and $0.03060990665$. This intervention compares input structure under a fixed channel; its error ratio is not a time or circuit-depth speedup.

Independently propagated block states agree with the Fig.~3 trajectory within $8.2\times10^{-16}$ entrywise. The independently solved reference differs by $2.4\times10^{-16}$ in Frobenius norm. Spectral reconstruction, the projector eigenrelations, projected channel evolution, and an independent sparse-exponential replay of the final block agree within $3\times10^{-16}$. Input infidelities match within $2\times10^{-13}$. The amplitude--direction identity and analytic optimal times are checked independently against direct norm evaluation and scalar minimization. Data, full density matrices, spectral projections, plotted coordinates, validation records, and a script reproducing the calculation and figure are provided in the accompanying data and code.

\subsection{Last-layer alignment and reachability}
The exact affine decomposition gives $X_P=A_P+\Delta_P$. For nonzero terms define
\begin{equation}
 \cos\vartheta_P=\frac{\operatorname{Re}\tr(A_P^\dagger\Delta_P)}{\norm{A_P}_F\norm{\Delta_P}_F}.
 \label{eq:smaligncos}
\end{equation}
The squared endpoint error is $\norm{A_P}_F^2+\norm{\Delta_P}_F^2+2\norm{A_P}_F\norm{\Delta_P}_F\cos\vartheta_P$. A negative cosine identifies cancellation, but small endpoint error also requires compatible magnitudes. When either term vanishes, the norm identity still holds without defining a cosine.

To identify dissipative directions, let $\mathcal B_P=e^{\tau_P\Dcal}$ and restrict it to the real Hilbert space $V_0$ of traceless Hermitian operators with Frobenius inner product. Trace preservation makes this space invariant. Its singular-vector decomposition is $\mathcal B_P(V_j)=s_jW_j$. Setting $\delta_j=\langle W_j,\Delta_P\rangle_F$ gives
\begin{equation}
 \norm{X_P}_F^2=\sum_j\left|s_j\langle V_j,\mathcal U_P(X_{P-1})\rangle_F+\delta_j\right|^2.
 \label{eq:smdirectional}
\end{equation}
Here $\mathcal U_P(X)=U_PXU_P^\dagger$. Coherent control and earlier layers determine which incoming components face small $s_j$ and how the remaining terms balance $\delta_j$. Unitary right composition preserves the singular values on $V_0$, so coherent control changes their input orientation, not the fixed-exposure singular-value spectrum. These directions are generally different from the target-Liouvillian eigenmodes. A nonunital many-qubit channel need not contract the Frobenius norm in every direction; trace-norm contractivity on Hermitian deviations remains valid. The decomposition does not presume $s_j\leq1$.

For finite exposures, products of matrix exponentials are invertible. With $\omega_P=\mathcal B_P^{-1}(\rho_{\rm ss})$ the exact identity is $X_P=\mathcal B_P[\mathcal U_P(\rho_{P-1})-\omega_P]$. The inverse preserves trace and Hermiticity but generally not positivity. Exact preparation by the fixed last channel requires $\omega_P$ to be a density matrix and to lie in the reachable coherent orbit of the preceding state. Even arbitrary unitaries preserve its spectrum; the two available coherent groups impose further restrictions.

For a single damped qubit, write $\rho_{\rm ss}=(I+\mathbf r_{\rm ss}\cdot\boldsymbol\sigma)/2$. Loss toward $\ket1$ sends $(x,y,z)$ to $(\sqrt\eta x,\sqrt\eta y,\eta z-(1-\eta))$. The required input vector is therefore
\begin{equation}
 \mathbf r_{\rm pre}=\left(\frac{r_{{\rm ss},x}}{\sqrt\eta},\frac{r_{{\rm ss},y}}{\sqrt\eta},\frac{r_{{\rm ss},z}+1-\eta}{\eta}\right).
 \label{eq:smpreimagebloch}
\end{equation}
It is physical exactly when $|\mathbf r_{\rm pre}|\leq1$. When physical, coherent rotations can match its orientation if the preceding state has the required Bloch-vector length; previous dissipation can adjust that length. A large final exposure can instead make this target input unphysical. This illustrates why last-layer alignment is a joint constraint on the learned path and the final channel, not a prescription to place all dissipation at the end. The Heisenberg obstruction below makes this distinction explicit for the actual boundary-driven controls.

\subsection{An exact obstruction to terminal-only boundary dissipation}
For $\rho_{\rm term}=e^{\tau\Dcal_N}\mathcal U_{\rm coh}(\rho_{0,N})$, let $\mathcal U_{\rm coh}$ be any sequence of the two allowed coherent groups. Both preserve the all-down density matrix: $H_1\ket1^{\otimes N}=0$ and $H_2\ket1^{\otimes N}=(N-1)\ket1^{\otimes N}$. The terminal reservoir operation acts only at the boundaries, leaving every interior spin down.

Let $\Pi_\downarrow$ project onto states with all interior spins down and arbitrary endpoints. Then $\tr(\Pi_\downarrow\rho_{\rm term})=1$. Reflection combined with global spin flip preserves the Heisenberg generator, so its unique target has equal all-down and all-up interior probabilities. These events are disjoint for $N\geq3$, giving $p_\downarrow=\tr(\Pi_\downarrow\rho_{\rm ss})\leq1/2$. Measurement of $\{\Pi_\downarrow,I-\Pi_\downarrow\}$ yields \mcite{jozsa1994,fuchs1999}
\begin{equation}
 D_{\rm tr}(\rho_{\rm term},\rho_{\rm ss})\geq1-p_\downarrow\geq\tfrac12,
 \quad F(\rho_{\rm term},\rho_{\rm ss})\leq p_\downarrow\leq\tfrac12.
 \label{eq:terminalbound}
\end{equation}
The bound uses the specified input and generators. Interleaved boundary pumping and coherent transport evade this obstruction; controls that directly flip bulk spins would change its premises.

\subsection{Boundary injection constrains arbitrarily large frozen transfer}
A separate limitation follows from excitation conservation. Define $\hat n=\sum_j(I+Z_j)/2$, which counts up spins. Both coherent groups commute with $\hat n$. During a boundary dissipative interval,
\begin{equation}
 \frac{d\langle\hat n\rangle}{d\tau}
 =1.2-1.2(\langle n_1\rangle+\langle n_N\rangle)\leq1.2.
\end{equation}
The initial expectation is zero. Therefore any LVA protocol with total dissipative exposure $T_D=\sum_\ell\tau_\ell$ satisfies $\langle\hat n\rangle_{\rm out}\leq1.2T_D$. The target symmetry gives $\langle\hat n\rangle_{\rm ss}=N/2$. Since $0\leq\hat n/N\leq I$, the variational characterization of trace distance yields
\begin{equation}
 D_{\rm tr}(\rho_{\rm out},\rho_{\rm ss})\geq
 \max\left(0,\frac12-\frac{1.2T_D}{N}\right).
 \label{eq:injection}
\end{equation}
Thus a frozen finite-exposure protocol cannot retain arbitrarily high global accuracy as $N\to\infty$. Depth growth, longer reservoir exposure, a different initial population, or additional number-changing controls change the premise. This model-specific bound is compatible with the observed finite-size block-count advantage.

The bulk-damped TFIM does not obey this boundary-only injection restriction. Neither model is classified as thermodynamically gapped or gapless here: that would require a separate Liouvillian scaling analysis, with attention to mode overlaps and nonnormality \mcite{znidaric2015,minganti2018,kastoryanotemme2013}.

\section{Finite-angle control response and correction}
\label{sec:response}
Let $\mathbf r_N(\Theta)$ concatenate the real and imaginary matrix components of each $R_{A_i}=\tr_{\bar A_i}\Lcal_N(\rho_\Theta)$, divided by $\sqrt{n_w}$, where $n_w=N-r+1$. Write $C_N(\Theta)=\norm{\mathbf r_N(\Theta)}^2=\mathcal C_{\rm loc}^{(r)}$ and define $J_N=\partial_\Theta\mathbf r_N$ and $K_N=J_N^TJ_N$. These matrices describe the response of the local stationarity residual to the physical controls. The same parameter units and site-sharing convention are used at every size.

\subsection{Restricted response strength and residual alignment}
Choose a source-selected orthonormal matrix $V$ in parameter space and set $A_N=J_NV$, $G_N=A_N^TA_N$. For $G_N$ positive definite and $\mathbf r_N\ne0$, let $\mu_N=\lambda_{\min}(G_N)$ and $\chi_N=\norm{\Pi_{\operatorname{ran}A_N}\mathbf r_N}^2/\norm{\mathbf r_N}^2$. Writing a singular-value decomposition $A_N=U\Sigma W^T$ gives
\begin{align}
 \norm{\nabla_x C_N(\Theta+Vx)|_{x=0}}^2
 &=4\norm{\Sigma U^T\mathbf r_N}^2
 \geq4\mu_N\chi_N C_N,\\
 x_*&=-A_N^+\mathbf r_N,\quad
 \min_x\norm{\mathbf r_N+A_Nx}^2=(1-\chi_N)C_N,\\
 \norm{x_*}&\leq\sqrt{\chi_N C_N/\mu_N}.
\end{align}
These are standard local least-squares identities applied to the physical residual. Sensitivity alone does not imply useful descent: $\chi_N$ can vanish. Nor is the unconstrained least-squares projection a guaranteed nonlinear improvement. If the restricted Jacobian has Lipschitz constant $L_J$ in an admissible neighborhood, then
\begin{equation}
 \norm{\mathbf r_N(\Theta+Vx)-\mathbf r_N-A_Nx}
 \leq\tfrac12L_J\norm{x}^2.
\end{equation}
A sufficiently small feasible step along a nonzero projected negative gradient therefore decreases the cost. Uniform response, alignment, and smoothness bounds over a neighborhood would be needed for a local convergence rate throughout that neighborhood. No smallness of the learned angles is assumed.

For the same $V$ at source and target, if
\begin{equation}
 \delta_N=\norm{G_s^{-1/2}G_NG_s^{-1/2}-I}_2<1,
 \quad\text{then}\qquad G_N\succeq(1-\delta_N)G_s.
\end{equation}
The minimum generalized eigenvalue is $\lambda_{\min}(G_s^{-1/2}G_NG_s^{-1/2})$; it measures retained response directly, without requiring the displayed closeness condition to hold. The response geometry concerns a chosen physical parameterization, as with other geometry-based variational methods \mcite{stokes2020}. It is not a claim that $J^TJ$ is a newly defined quantum information metric.

\subsection{Feasible directions and numerical protocol}
The checkpoints are the five-spin infidelity-trained protocols used in main Figs.~3 and~5: TFIM at $P=7$ and Heisenberg at $P=20$. Their parameters are held fixed at each size $N=5,\ldots,12$ before the diagnostic correction. Coordinates within $10^{-7}$ of either source box bound are held fixed, leaving 20 and 53 free directions, respectively. This confines corrections to a feasible face and can omit useful directions that activate an idle dissipative channel. The leading $k=3,6,10$ source eigenvectors of the free-coordinate $K_s$ define $V$; no target fidelity enters this selection. SVD projections retain singular values larger than $\max(10^{-9}s_{\max},10^{-12})$.

For the one-step diagnostic, constrained linearized least squares is solved within $\norm{\delta\Theta}_2\leq0.1$ and the physical parameter bounds, using SLSQP with a 200-iteration cap and tolerance $10^{-14}$. A line search compares six step lengths, $1,1/2,1/4,1/8,1/16,1/32$, using only the nonlinear local cost; the unchanged parameters remain eligible. The full target reference is consulted afterward for fidelity and the global residual. All 32 constrained subproblems, using either six source directions or all free directions at the 16 checkpoints, return successful solver flags. This one-step experiment is a local response diagnostic, not a replacement for the 1000-iteration target refinements in main Figs.~5 and~6.

Forward tangent responses agree with independent adjoint gradients within $1.0\times10^{-15}$ across the tested checkpoints. Central finite differences test every coordinate at $N=3$ using two-site residuals, and three coordinates at each of $N=10,11,12$; the maximum Jacobian discrepancy is $1.69\times10^{-10}$. Independent propagation reproduces the saved frozen infidelities within $4.0\times10^{-13}$. The dense and GPU representations agree at $N=8$ to $2.8\times10^{-13}$ in entries of $J_N^TJ_N$. Tangent operators and excitation sectors are never normalized separately. Full response matrices, corrected controls, line-search records, solver flags, and input hashes are provided in the accompanying data and code.

\begin{table}[ht]
\centering\small
\caption{Response diagnostics for the infidelity-trained protocols. $\eta_N=\lambda_{\min}(G_s^{-1/2}G_NG_s^{-1/2})$ uses six source-selected directions; $\chi_6$ and $\chi_{\rm all}$ are linearized correctable residual fractions. Cost reduction and final infidelity refer to one step using all free directions. The $N=5$ rows are source diagnostics.}\label{tab:response}
\begin{tabular}{lrrrrrr}\toprule
Model & $N$ & $\eta_N$ & $\chi_6$ & $\chi_{\rm all}$ & Cost reduction & Infidelity after step\\\midrule
TFIM & 5 & 1.0000 & 0.2377 & 0.9052 & 60.58\% & 0.002364 \\
TFIM & 6 & 0.7203 & 0.2445 & 0.9290 & 62.39\% & 0.003196 \\
TFIM & 7 & 0.6546 & 0.3020 & 0.9349 & 67.03\% & 0.003941 \\
TFIM & 8 & 0.5748 & 0.3343 & 0.9223 & 69.65\% & 0.004861 \\
TFIM & 9 & 0.5053 & 0.3621 & 0.9213 & 72.26\% & 0.006006 \\
TFIM & 10 & 0.4510 & 0.3874 & 0.9223 & 73.74\% & 0.007348 \\
TFIM & 11 & 0.4079 & 0.4104 & 0.9220 & 74.71\% & 0.008844 \\
TFIM & 12 & 0.3726 & 0.4317 & 0.9229 & 75.56\% & 0.010429 \\
Heisenberg & 5 & 1.0000 & 0.0497 & 0.8719 & 58.91\% & 0.000355 \\
Heisenberg & 6 & 0.3020 & 0.2405 & 0.9907 & 66.00\% & 0.011739 \\
Heisenberg & 7 & 0.4047 & 0.2297 & 0.9659 & 64.29\% & 0.036549 \\
Heisenberg & 8 & 0.2831 & 0.2439 & 0.9335 & 61.15\% & 0.067975 \\
Heisenberg & 9 & 0.2565 & 0.2023 & 0.9121 & 52.48\% & 0.121331 \\
Heisenberg & 10 & 0.2264 & 0.2470 & 0.9177 & 57.71\% & 0.199410 \\
Heisenberg & 11 & 0.2395 & 0.1413 & 0.8923 & 46.30\% & 0.319168 \\
Heisenberg & 12 & 0.2536 & 0.1576 & 0.8846 & 46.72\% & 0.449168 \\
\bottomrule\end{tabular}
\end{table}

Table~\ref{tab:response} separates retained response strength from error alignment and actual correction. At $N=12$, the minimum generalized response eigenvalues are 0.3726 for TFIM and 0.2536 for Heisenberg, while the six-direction correctable fractions are 0.4317 and 0.1576. All free directions address larger linearized fractions, 0.9229 and 0.8846. The constrained all-free step lowers the local cost by 75.56\% and 46.72\%, with infidelities changing from 0.022361 to 0.010429 and from 0.457971 to 0.449168, respectively. Local improvement does not guarantee global improvement: at $N=9$, the Heisenberg local cost decreases by 52.48\% while infidelity changes from 0.120130 to 0.121331. Both source checkpoints also have larger infidelity after the local-cost correction. Thus retained response and substantial stationarity correction do not certify global state accuracy or an optimal low-dimensional correction basis.

The sufficient closeness condition $\delta_N<1$ holds at all tested TFIM target sizes but fails for Heisenberg at $N=6,7$, where $\delta_N$ is 1.3207 and 1.2125, respectively. The positive minimum generalized eigenvalues still quantify retained response at those sizes. At $N=12$, the median correctable fractions over 20 random six-dimensional subspaces in the same free parameter space are 0.5104 and 0.1688, respectively, exceeding the selected leading-six fractions in both models. Strong source response therefore does not by itself select the best residual-correction directions.

\subsection{Spatial sharing and exact fixed-depth locality in TFIM}
Onsite rotations and dissipative channels preserve observable support. Since the nearest-neighbor $ZZ$ terms commute, each complete $ZZ$ block enlarges the support by at most one site on either side. After the target adjoint generator has acted on an $r$-site Pauli observable, a conservative backward radius for $P$ layers is $R=P+1$. This support bound is independent of the finite angle values. The homogeneous product input and site-shared parameters imply that all interior windows have the same residual function $f_{\rm b}(\Theta)$ and Jacobian $D f_{\rm b}(\Theta)$.

If $n_w=N-r+1>2R$, there are $n_w-2R$ identical interior contributions and $2R$ fixed boundary contributions $f_a(\Theta)$. Consequently,
\begin{align}
 C_N&=\norm{f_{\rm b}}^2+\frac{1}{n_w}
       \sum_{a=1}^{2R}\bigl(\norm{f_a}^2-\norm{f_{\rm b}}^2\bigr)
       =c_{\rm b}+\frac{\beta}{n_w},\\
 K_N&=(Df_{\rm b})^TDf_{\rm b}+\frac{1}{n_w}
       \sum_{a=1}^{2R}\bigl[(Df_a)^TDf_a-(Df_{\rm b})^TDf_{\rm b}\bigr]
       =K_{\rm b}+\frac{K_{\partial}}{n_w}.
\end{align}
The correction matrix $K_{\partial}$ need not be positive. These are exact functional identities, so derivatives can be taken before or after separating bulk and edges. Each gradient component has form $a(\Theta)+b(\Theta)/n_w$. For an $N$-independent parameter-neighborhood distribution, its second moment is a polynomial in $1/n_w$, not a new exponential-in-$N$ factor. A direction may be bulk-responsive, boundary-responsive, or identically inactive. The coefficients can still be very small at large depth, and alignment with the residual is not guaranteed.

Spatial sharing is essential to comparing derivatives in a common finite-dimensional parameter space. With independent site parameters, new columns of $J_N$ would be introduced as $N$ grows, and the matrix identity above would no longer have the same meaning. The result applies to the local cost, not the full HS residual or global fidelity. It also does not prove five-spin transfer when $N_s$ fails to enclose the conservative causal region. For generic local channels, approximate propagation bounds provide a possible extension \mcite{poulin2010,barthel2012}; the exact grouped Heisenberg exchange operation requires that more general treatment.

Two untrained finite-angle schedules independently check the bulk/edge identities for $C_N$, $K_N$, and $J_N^T\mathbf r_N$. Their layer parameters $(a,b,\tau)$ are $(1.1,-0.7,0.45)$ for $(P,r)=(1,2)$ and $[(1.1,-0.7,0.45),(0.9,0.4,0.6)]$ for $(P,r)=(2,1)$. The tested sizes are $N=6,\ldots,12$ and $N=7,\ldots,12$, respectively. In each case, the first two sizes determine the bulk and boundary coefficients; the remaining sizes are independently propagated and compared with the predicted affine dependence on $1/(N-r+1)$. The maximum absolute componentwise discrepancies across all three quantities are $7.82\times10^{-14}$ and $9.55\times10^{-15}$, respectively. The starting sizes satisfy $N-r+1>2(P+1)$, requiring $N\geq6$ for $(P,r)=(1,2)$ and $N\geq7$ for $(P,r)=(2,1)$. Table~\ref{tab:bulkchecks} lists all sizes; the first two sizes determine the coefficients, and each remaining size provides an independent check.

\begin{table}[ht]
\centering\small
\caption{Fixed-depth TFIM bulk/edge identity checks. Errors are absolute differences for $C_N$ and maximum absolute componentwise differences for $K_N$ and $J_N^T\mathbf r_N$. Dashes mark the first two sizes of each sequence, which determine the coefficients; all remaining entries are independent prediction checks.}\label{tab:bulkchecks}
\begin{tabular}{rrrrrr}\toprule
$P$ & $r$ & $N$ & Error in $C_N$ & Error in $K_N$ & Error in $J_N^T\mathbf r_N$\\\midrule
1 & 2 & 6 & -- & -- & -- \\
1 & 2 & 7 & -- & -- & -- \\
1 & 2 & 8 & $1.55\times10^{-15}$ & $1.78\times10^{-14}$ & $4.00\times10^{-15}$ \\
1 & 2 & 9 & $1.78\times10^{-15}$ & $2.66\times10^{-14}$ & $2.66\times10^{-15}$ \\
1 & 2 & 10 & $9.33\times10^{-15}$ & $7.11\times10^{-14}$ & $1.15\times10^{-14}$ \\
1 & 2 & 11 & $9.99\times10^{-15}$ & $7.28\times10^{-14}$ & $1.11\times10^{-14}$ \\
1 & 2 & 12 & $9.77\times10^{-15}$ & $7.82\times10^{-14}$ & $1.42\times10^{-14}$ \\
2 & 1 & 7 & -- & -- & -- \\
2 & 1 & 8 & -- & -- & -- \\
2 & 1 & 9 & $2.22\times10^{-16}$ & $1.11\times10^{-15}$ & $2.22\times10^{-16}$ \\
2 & 1 & 10 & $1.89\times10^{-15}$ & $6.66\times10^{-15}$ & $3.33\times10^{-15}$ \\
2 & 1 & 11 & $2.11\times10^{-15}$ & $7.33\times10^{-15}$ & $3.33\times10^{-15}$ \\
2 & 1 & 12 & $2.44\times10^{-15}$ & $9.55\times10^{-15}$ & $4.22\times10^{-15}$ \\
\bottomrule\end{tabular}
\end{table}

These tests establish the algebraic identity, not the optimization performance of the untrained schedules. The identity depends on the physical channel structure and fixed depth, not on the source-training loss. The conditional response analysis differs from a short-total-evolution-time HVA guarantee \mcite{park2024} and is closer in purpose to studying useful neighborhoods of transferred solutions \mcite{mele2022,puig2025}.

\section{Residual conditioning, global accuracy, and reproducibility}
\label{sec:limits}
Let $V_0$ be the real vector space of traceless Hermitian operators. For a unique steady state, define the smallest singular value of the generator restricted to this space,
\begin{equation}
 \sigma_{*,N}=\inf_{X\in V_0\setminus\{0\}}
 \frac{\norm{\Lcal_N(X)}_F}{\norm{X}_F}>0.
\end{equation}
For a physical state $\rho$ and $d_N=2^N$, linearity and the trace/Frobenius norm inequality give
\begin{equation}
 \norm{\rho-\rho_{\rm ss}}_F\leq\frac{\sqrt{\mathcal C_{\rm res}}}{\sigma_{*,N}},
 \quad D_{\rm tr}(\rho,\rho_{\rm ss})\leq
 \frac{\sqrt{d_N\mathcal C_{\rm res}}}{2\sigma_{*,N}}.
 \label{eq:conditioning}
\end{equation}
The real-part Liouvillian gap is not a replacement for $\sigma_{*,N}$ in a nonnormal problem. Nor does closing a real-part gap necessarily make the modulus of an eigenvalue approach zero. These are distinct spectral and conditioning questions.

A simple example separates small residual from state accuracy. For $N$ independently damped spins with rate $\gamma$, let $\rho=I/2^N$ and $\rho_{\rm ss}=(\ket1\bra1)^{\otimes N}$. In the conventions of this work,
\begin{equation}
 \Lcal_N(I/2^N)=-\frac{\gamma}{2^N}\sum_j Z_j,
 \quad\mathcal C_{\rm res}=\frac{N\gamma^2}{2^N},
 \quad D_{\rm tr}=1-2^{-N}.
\end{equation}
The raw residual vanishes exponentially although the global error approaches one; every single-site state also remains wrong. Dividing by the purity removes this particular dilution, but changes the training objective and does not remove the conditioning factor in Eq.~\eqref{eq:conditioning}. The raw residual, local stationarity, and global fidelity are therefore distinct quantities. Source training directly minimizes infidelity, while target refinement minimizes the local stationarity residual.

A size-independent dissipative gap also does not imply size-independent global preparation time. Starting all spins excited under the same independent damping, the squared fidelity is $F_N(t)=(1-e^{-\gamma t})^N$. For small global infidelity $\epsilon$,
\begin{equation}
 t_\epsilon=-\gamma^{-1}\log\bigl[1-(1-\epsilon)^{1/N}\bigr]
 \simeq\gamma^{-1}\log(N/\epsilon).
\end{equation}
This elementary example is consistent with known quantum mixing and cutoff phenomena \mcite{cutoff2012,kastoryanotemme2013}. It motivates testing transferred acceleration rather than inferring it from a gap or a large system size. Conversely, local observable accuracy can remain stable while global infidelity grows. Neither local stationarity nor a global norm proxy should be silently substituted for the physical accuracy used in a resource ratio.

\paragraph*{Reproducibility.}
The accompanying data and code contain source and target parameters, objective histories, baseline time scans, reference-state checks, modal trajectories, response matrices, and the records underlying all six main figures. Source training minimizes infidelity; target refinement minimizes local stationarity residuals. Both use 1000-iteration caps per optimizer call in the transfer and resource tests. Numerical checks cover trace preservation, channel/adjoint consistency, dense-versus-sector evolution, finite-difference gradients, exact second-order inclusion, and finite-angle locality. Solver settings, reference-state checks, dependency hashes, and source-evaluation totals are also provided. The accompanying README maps the figures and supplementary analyses to their data files and reproduction scripts. Tensor-network and trajectory approaches offer additional classical routes to larger systems where justified \mcite{verstraete2004,zwolak2004,werner2016,weimer2021}.

\bibliographystyle{lva-preserve}
\bibliography{bibliography/supplement}